# Learning qBIC Resonances across Metasurface Families in Dielectric Fourier Space

Shuangteng Lei[1,2,†], Li Yu[1,2,†,*], Tianxin Li[1,2,*], Wei Lu[1,2,3,*]

1 State Key Laboratory of Infrared Physics, Shanghai Institute of Technical Physics, Chinese Academy of Sciences, Shanghai 200083, China;

2 University of Chinese Academy of Sciences, Beijing 100049, China

3 School of Physical Science and Technology, ShanghaiTech University, Shanghai 201210, China;

†These authors contributed equally to this work: Shuangteng Lei and Li Yu.

*Correspondence: Li Yu (yuli@mail.sitp.ac.cn), Tianxin Li (txli@mail.sitp.ac.cn) and Wei Lu (luwei@mail.sitp.ac.cn).

**Abstract**

Bound states in the continuum (BIC) metasurfaces are typically described by geometry-specific parameters, hindering cross-geometry comparison, while ultranarrow quasi-BIC (qBIC) features are easily diluted in full-spectrum learning. Here, we map 2,015 samples from seven dielectric metasurface families onto a shared reciprocal-lattice grid. Two fixed low-order Fourier descriptors capture within-branch resonance shifts, with mean $R^2$ values of 0.871–0.999. Field-level analysis of two representative branches shows that these shifts are consistent with the Maxwell–Fourier perturbation picture. A five-channel Fourier-space backbone models the broadband spectrum. A local resonance expert uses complex Fourier coefficients to parameterize the qBIC resonance through a differentiable Fano layer. The expert reduces resonance-position mean absolute error (MAE) from 3.2 to 0.95 nm and reduces the minimum-transmission MAE by a factor of 14 on a geometry-blocked test set. The same coordinate supports spectrum-to-structure reconstruction.

Neural-network surrogates can replace repeated full-wave simulations for response prediction and inverse design[1–5]. Many models use family-specific parameters, such as radius, gaps or rotation angles[1,4,6]. Two difficulties remain when learning narrow resonances across different metasurface families. Models based on family-specific parameters lack a shared structural coordinate. A narrow resonance occupies only a small fraction of the sampled wavelengths and contributes little to a loss averaged over the full spectrum. Neural networks tend to learn low-frequency components first[7]. This tendency further motivates treating sharp features explicitly.

These difficulties are particularly relevant to metasurfaces supporting photonic bound states in the continuum (BICs). A photonic BIC is a nonradiating mode within the radiation continuum[8,9]. Symmetry breaking or finite size can introduce radiative leakage and produce a quasi-BIC (qBIC)[8,10]. The qBIC can appear as a narrow Fano resonance[11]. Such resonances support low-threshold lasing[12], molecular fingerprinting[13], biosensing[14] and nonlinear optics[15]. The real and imaginary parts of the complex eigenfrequency specify the modal frequency and decay, which together determine the quality factor Q. Symmetry, interference and polarization topology explain BIC formation and radiative leakage[10,16–18]. Yet many models still learn qBIC spectra from family-specific structural parameters. Resonance-aware strategies have improved the prediction and inverse design of high-Q resonances[19,20]. Joint learning across distinct families also needs a shared structural representation.

Dielectric Fourier coefficients provide a common description of periodic structures. In the plane-wave formulation of Maxwell's equations, $\varepsilon_{\mathbf{G}-\mathbf{G}'}$ couples Bloch harmonics $\mathbf{G}$ and $\mathbf{G}'$ [21,22]. Mode-expansion analyses explain radiation cancellation in photonic-crystal-slab BICs using this framework[23]. Individual permittivity harmonics have also been used to control band flips, radiative coupling and diffraction[24–26]. Fourier-designed surface profiles enable wavefront generation and sensing in compact pixels[27]. Structures sharing a lattice, origin and orientation occupy the same reciprocal-lattice grid, allowing comparison without matching their real-space parameters. Given the close correspondence between optical diffraction and Fourier transforms, these coordinates may reveal shared structure–spectrum relationships across geometries. Fourier neural operators learn global convolutions in a truncated spectral basis28. Here, Fourier coordinates reorganize structural pixels into diffraction-related spatial harmonics. Selecting dominant low-order complex coefficients for the local expert implicitly redistributes learning weight towards the coupling amplitudes and phases governing the target resonance.

Here, we use dielectric Fourier space to compare and model qBIC resonances in seven families of silicon metasurfaces (Fig. 1). First-order perturbation theory and feature-importance analysis guide our choice of two low-order descriptors. We keep their definitions frozen across families. The descriptors track resonance shifts within continuous parameter branches. Field-overlap calculations support the predicted linear trends for two representative branches. A Fourier-space backbone predicts the broadband response. A local expert uses complex Fourier coefficients to predict the resonance centre, quality factor, depth and Fano asymmetry parameter. A differentiable Fano layer converts these parameters into a resonance lineshape. Adding the expert reduces the resonance-position mean absolute error from 3.2 to 0.95 nm on the geometry-blocked test set. Predicting Fourier coefficients also improves spectrum-to-structure reconstruction over direct pixel prediction. The model combines a physical input representation with a Fano output layer. These choices can complement methods based on governing equations in the training loss[29,30].

## Results

### Dielectric Fourier coefficients as a shared structural coordinate

The seven families—Asymmetric Cross (AC), Asymmetric Disks (AD), Circular Split Ring (CSR), Disk with an Asymmetric Nanohole (DAN), Parallel Bars with an x-offset (PX), Stepped Rectangle (SR) and Twin Yawed Ellipses (TY) (Fig. 2a)—are silicon motifs on a square lattice of period 1,320 nm, adapted from published qBIC designs[10,31–36]. Each motif is controlled by two geometric parameters, $P_1$ and $P_2$. All families share the lattice, coordinate origin and orientation, film thickness and normal-incidence excitation (Methods and Supplementary Tables S1 and S2). Scanning one parameter with the other fixed produces a continuous qBIC branch (Fig. 2c and Supplementary Figs. S1–S7). These parameters cannot be compared directly across families. For example, a radius offset in AD has no counterpart in the rotation angle of TY. The centred two-dimensional fast Fourier transform (FFT) of each unit cell's permittivity places all 2,015 structures on one reciprocal-lattice grid (Fig. 1a,b). Unlike a far-field angular spectrum, these coefficients follow from the geometry alone and require no electromagnetic solution.

For an in-plane periodic structure, the permittivity is expanded as

$$\varepsilon(\boldsymbol{\rho}, z) = \sum_{\mathbf{q}} \varepsilon_{\mathbf{q}}(z) e^{i\mathbf{q}\cdot\boldsymbol{\rho}}, \qquad \boldsymbol{\rho} = (x, y). \tag{1}$$

where $\mathbf{q}$ denotes a reciprocal-lattice vector and the numerical FFT samples $\varepsilon_{\mathbf{q}}$. Expanding the Bloch field in plane-wave harmonics $\mathbf{G}$ and substituting into Maxwell's equations gives[21,22]

$$\mathbf{L_G E_G} = \frac{\omega^2}{c^2} \sum_{\mathbf{G}'} \varepsilon_{\mathbf{G}-\mathbf{G}'} \mathbf{E}_{\mathbf{G}'}, \tag{2}$$

so that the dielectric part of the eigenvalue matrix is

$$B_{\mathbf{G},\mathbf{G}'} = \varepsilon_{\mathbf{G}-\mathbf{G}'}. \tag{3}$$

A structural Fourier component $\varepsilon_{\mathbf{q}}$ therefore couples every pair of harmonics whose wavevectors differ by $\mathbf{q}$. This is the photonic counterpart of the nearly-free-electron model[37]. In that model, the Fourier component of the crystal potential couples plane waves separated by a reciprocal-lattice vector and sets the gap at the zone boundary. A change in geometry alters these couplings and shifts the complex eigenfrequency of the mode.

For weak leakage the mode can be treated as approximately Hermitian. First-order perturbation of the generalized eigenproblem, with the in-plane integral transformed to Fourier space, gives the shift of the real eigenfrequency[38] (Supplementary Note 2)

$$\Delta\omega_i \simeq -\frac{\omega_0}{2T}\,\mathrm{Re}\sum_{\mathbf{q}} \Delta\,\varepsilon_{i,\mathbf{q}} W_{\mathbf{q}}^{(r)}, \tag{4}$$

where

$$W_{\mathbf{q}}^{(r)} = \sum_{\mathbf{G}} \int dz\, \mathbf{E}_{r,\mathbf{G}}^{*}(z)\cdot \mathbf{E}_{r,\mathbf{G}-\mathbf{q}}(z), \qquad T = \int dV\, \mathbf{E}_r^{*}\cdot \varepsilon_r \mathbf{E}_r. \tag{5}$$

$W_{\mathbf{q}}^{(r)}$ is the overlap between harmonics of the reference mode $r$ separated by $\mathbf{q}$, and $T$ is the modal normalization. The resonance shift sums structural Fourier changes weighted by these harmonic overlaps. Two limitations apply. First-order perturbation in $\Delta\varepsilon$ is not exact for moving boundaries with high refractive-index contrast. We therefore use Eq. (4) to identify relevant wavevectors and test its trends numerically. The weights $W_{\mathbf{q}}$ also depend on the mode. Thus, the same $\Delta\varepsilon_{\mathbf{q}}$ can shift resonances in different families by different amounts or in opposite directions.

For the near-Γ qBICs studied here, most modal weight lies in low-order harmonics, so the largest overlaps occur at small nonzero $\boldsymbol{q}$. This motivates two representations. Scalar descriptors summarize low-order coefficients, while a local 9 × 9 complex block retains their relative amplitudes and phases for the resonance expert (Fig. 1c).

**Fixed low-order Fourier descriptors capture dominant resonance shifts**

To locate the wavevectors associated with resonance shifts, we trained two ExtraTrees regressors on the 67 × 67 Fourier pixels for resonance position and ln $Q$. Both importance maps concentrate near Γ and are approximately mirror-symmetric, but their directional preference differs between families (Fig. 3d). These characteristics motivated two complementary low-dimensional descriptors (Fourier channels): a near-Γ descriptor for the dominant low-order Fourier content and a symmetry-aware anti-diagonal descriptor for the directional modulation in the importance maps. We defined two scalar descriptors from the regions identified by ExtraTrees and kept their definitions frozen for all families, scan directions and slices.

The first descriptor, $C_\Gamma$, averages the log-amplitude of the 48 nonzero harmonics in the central 7 × 7 block:

$$C_\Gamma = \frac{1}{48} \sum_{\substack{|m|\leq 3, |n|\leq 3 \\ (m,n)\neq(0,0)}} \ln\left(1 + |F_{mn}|\right). \tag{6}$$

The second descriptor, $C_D$, is the anti-diagonal low-order modulation power:

$$C_D = \sum_{(m,n)\in\mathcal{D}} |F_{mn}|^2, \qquad \mathcal{D} = (1,-1), (-1,1), (2,-2), (-2,2). \tag{7}$$

Because the permittivity is real, its spectrum is Hermitian-symmetric ($F_{-m,-n} = F^*_{mn}$). The four points in $\mathcal{D}$ therefore carry two independent complex amplitudes. For each family we used the descriptor with the higher median within-branch $R^2$: $C_\Gamma$ for CSR and PX and $C_D$ for the other five (Supplementary Table S3).

Along the 17-point branches, the selected descriptor is linear in the resonance position. The mean within-branch $R^2$ is 0.984–0.999 for AC, AD, CSR, DAN, SR and TY and 0.871 for PX (Figs. 2d and 3b,f; all branches in Supplementary Figs. S8–S14). PX spans the widest parameter range. The harmonic content of the mode changes appreciably over this range, and a single amplitude-only scalar cannot follow such changes in $W_q$. Figure 3e overlays all families after each slice is affinely rescaled by its fitted offset and span. This rescaling removes the family-specific slopes. The collapse onto $y = x$ therefore reflects linearity within each branch, not a universal slope. The fits are in-sample and descriptive; out-of-sample prediction is evaluated with the neural models below.

The same descriptors describe ln $Q$ less uniformly. Within-branch $R^2$ exceeds 0.8 for AC, CSR, DAN and TY but is 0.670, 0.765 and 0.381 for AD, PX and SR (Fig. 3c,f). This follows from the form of the radiative loss. With $Q = \omega_r/\kappa$ and radiation-dominated loss, expanding the outgoing amplitudes to first order in Δε gives (Supplementary Note 2)

$$\ln Q \simeq C - \ln\left[\sum_{\mu}\left|d_{\mu,0} + \sum_{\mathrm{q}} R_{\mu\mathrm{q}}\,\Delta\varepsilon_{\mathrm{q}}\right|^2\right]. \tag{8}$$

where $d_{\mu,0}$ is the radiation amplitude of the reference state in open channel μ and $R_{\mu\mathbf{q}}$ is the overlap of wavevector **q** with that channel. In contrast to Eq. (4), Eq. (8) contains a coherent sum within each channel. Squaring it produces cross terms between wavevectors, so $Q$ depends on the relative phases of the Fourier coefficients as well as on their amplitudes. Amplitude-only scalars cannot capture this interference in general. The full lineshape requires the complex local representation.

**Field-level Maxwell–Fourier validation**

To test whether the fixed descriptors identify wavevectors that are physically relevant to the field-weighted resonance shift, we evaluated the field-overlap response on two continuous qBIC branches. The TY branch represents $C_D$, and the CSR branch represents $C_\Gamma$. For each branch, we selected a reference state r and constructed its Fourier-harmonic overlap from the three-dimensional complex driven field at its resonance wavelength. We then applied a two-dimensional x–y Fourier transform independently at each z layer:

$$E_G(z) = (1/N^2)FFTshift\big[FFT_{xy}E(x,y,z)\big] \quad (9)$$

Integrating over z gives the modal weight of each harmonic,

$$M_G = \int dz\, \Sigma_{\alpha=x,y,z}\big|E_{\alpha,G}(z)\big|^2 \quad (10)$$

The central 7 × 7 block holds 93.3% (TY) and 93.2% (CSR) of the total modal weight (Fig. 4b), and the concentration persists along both branches (Supplementary Fig. S15).

The Fourier-harmonic overlaps $W_q$ are calculated from Eq. (5). We further define

$$S_i = \mathrm{Re}\sum_{\mathbf{q}} \Delta\, \varepsilon_{i,\mathbf{q}} W_{\mathbf{q}}^{(r)}. \quad (11)$$

Eq. (4) predicts that $S_i$ is proportional to the resonance shift. Within each branch, $S_i$ computed using the descriptor wavevectors is linear in the resonance shift Δλ across the 17 states. The fits give $R^2$ = 0.996 for TY and 0.995 for CSR (Fig. 4d). Other reference states give $R^2$ = 0.986–0.997 (Supplementary Fig. S16). For TY, the four $C_D$ wavevectors give a tighter linear relation than the full sum ($R^2$ = 0.986–0.996 versus 0.959–0.968). This indicates that at first order the remaining wavevectors add scatter rather than trend. The descriptors can be interpreted as the first-order Maxwell–Fourier response compressed onto its dominant wavevectors.

**A local Fourier-space resonance expert reconstructs sharp qBIC resonances**

Although the fixed low-order descriptors track resonance-position shifts well, they describe Q less uniformly and cannot by themselves specify the full qBIC lineshape. The narrow resonance is also difficult to learn using a loss averaged over the full spectrum. the spectral loss is the mean error over $N_\lambda$ wavelength points,

$$\mathcal{L}_{\mathrm{spec}} = \frac{1}{N_\lambda}\sum_{i=1}^{N_\lambda}\big|T(\lambda_i) - \hat{T}(\lambda_i)\big| \quad (12)$$

a resonance only a few sampling points wide contributes little to the loss or its gradient. A model can reach a low spectral error while missing the resonance. However, the ExtraTrees importance maps above concentrate near Γ (Fig.3d), and the field analysis further supports the concentration of modal Fourier weight in the central low-order region (Fig.4b,c).

Guided by this localization, we introduce a local resonance expert that uses the central 9 × 9 Fourier block to learn the narrow resonance, alongside a broadband background branch (Fig. 1d–f). The backbone pools and concatenates convolutional features from the full input and its central 33 × 33 crop. A spectral multilayer perceptron (MLP) predicts 32 principal component analysis (PCA) coefficients for background reconstruction (Fig. 1d). The background branch is trained on the full spectrum, including the narrow resonance. It receives either the 67 × 67 real-space permittivity map (Real bg) or a five-channel Fourier tensor built from the complex coefficients $F_{(mn)}$ (K bg),

$$\mathbf{X}_K = [\mathrm{Re}F, \mathrm{Im}F, \log(1 + |F|), \sin\phi, \cos\phi], \qquad \phi = \arg(F). \tag{13}$$

The five channels are the real part, imaginary part, log-amplitude, phase sine and phase cosine. The expert pools features from the central 5 × 9 × 9 block for a peak MLP (Fig. 1c,e). Its outputs, the resonance centre $\lambda_0$, ln $Q$, depth $d$ and Fano asymmetry parameter $q_f$, enter a differentiable Fano layer to add the resonance to the background (Fig. 1f):

$$\hat{T}(\lambda) = \hat{T}_{\mathrm{bg}}(\lambda) + \Delta T_{\mathrm{Fano}}\left(\lambda; \lambda_0, Q, d, q_f\right) \tag{14}$$

Because the Fano layer is differentiable, the spectral error propagates directly to the four resonance parameters, giving the narrow feature its own gradient path. No independent label exists for $q_f$; it is learned through the spectral loss with a weak prior towards zero (Supplementary Note 5). Combining the two backbone inputs with the presence or absence of the expert gives four models: Real bg, K bg, Real + K peak and K + K peak. They differ only in these two factors.

All models use the same strict split. Source-grid geometry hashes and principal-scan parameter blocks are separated between training (1,556 samples), validation (238) and test (221) sets. All models share the PCA basis and evaluation protocol. At the 67 × 67 binary-mask resolution, 68 test masks coincide with a training mask; results for the remaining 153 are reported separately where relevant (Supplementary Fig. S17). Training and validation loss curves for the four models are shown by optimization stage in Supplementary Fig. S18.

Both background-only models follow the resonance shift along a continuous DAN slice but predict dips that are too broad and too shallow (Fig. 5a). Adding the expert brings the transmission minima onto the trajectory in the raw data. Over the test set and five random seeds, the expert lowers the resonance-position mean absolute error (MAE) from 3.000 to 0.998 nm on the real-space backbone and from 3.223 to 0.947 nm on the Fourier backbone. It reduces the minimum-transmission MAE by 91% and 93%, respectively (Table 1 and Fig. 5c). The ln $Q$ MAE falls from 0.542 to 0.289 and from 0.673 to 0.295. With the expert in place, the two backbones perform almost identically. The final resonance accuracy comes from the resonance branch (Supplementary Fig. S19). The gain is not uniform across families (Supplementary Fig. S20). It is large for AD, CSR, DAN, PX and SR. For AC and TY, the resonance features are comparatively broad and easier for the background models to learn. Conventional predictions already reach resonance-position errors on the order of 1 nm. The additional resonance expert therefore does not provide further improvement.

To separate the contribution of the Fano-parameterized branch from that of its input, we kept the real-space backbone and the expert architecture fixed and changed only the expert input (single seed; Supplementary Fig. S21). The resonance-position MAE is 2.05 nm with the full real-space map as expert input, 1.98 nm with the full Fourier tensor, 2.69 nm with a central 9 × 9 real-space crop, 2.55 nm with the two scalars $C_\Gamma$ and $C_D$, and 1.00 nm with the local 9 × 9 complex Fourier block. The decrease in MAE from 3.00 nm for the background-only model to 1.00 nm with the local complex Fourier block reflects the joint benefit of the Fano-parameterized branch and its input. This block retains low-order amplitude and phase information from the whole unit cell, whereas a 9 × 9 real-space crop captures only a small region. The two scalars suffice for the resonance position along a branch but discard the phase and directional information that the lineshape requires. Additional surrogate-model comparisons and representative spectra are provided in Supplementary Figs. S22 and S23. Supplementary Fig. S24 gives the FDTD, training and inference costs.

To evaluate whether the model can also predict spectra for unseen structures, we held out each of the seven structure families in turn and compared zero-sample and one-sample settings. In the zero-sample setting, the model was trained on the other families and showed substantial errors in resonance prediction for the held-out family. Adding just one target-family sample substantially improved full-spectrum prediction (Supplementary Fig. S25). These results suggest that the model learns relationships between structural changes and spectral responses across families. Limited target-family data can help transfer these relationships to a new family.

**Fourier coefficients as the output of inverse reconstruction**

The shared coordinate can also serve as the output of an inverse model. A one-dimensional convolutional encoder predicts the unit cell's complex Fourier coefficients from a spectrum in the target window. An inverse FFT then recovers the real-space structure. Family labels are used as auxiliary supervision during training and are not supplied at inference. We compare this Fourier route with direct decoding of the real-space mask and with joint supervision in both domains.

On the 221 test samples, the Fourier route reaches a mean foreground intersection-over-union (IoU) of 0.944, compared with 0.853 for direct real-space decoding. Joint supervision gives 0.926 (Fig. 5b and Supplementary Fig. S26). On the 153 test masks with no pixel-identical counterpart in training, the corresponding values are 0.927, 0.816 and 0.920. The Fourier and joint routes are statistically indistinguishable (paired IoU difference +0.019; 95% bootstrap interval −0.004 to 0.039). All inverse comparisons use a single seed. Predicting fourier coefficients appears to regularize the output towards smooth, connected shapes. This route also suppresses the fragmented artefacts of pixel-wise decoding (Supplementary Fig. S27).

## Discussion

The dielectric Fourier representation provides shared, physically meaningful coordinates for learning spectral responses across different geometries. The dielectric Fourier coefficients $\varepsilon_{\mathbf{q}}$ couple Bloch harmonics in the plane-wave Maxwell eigenproblem, allowing geometrically distinct structures to be compared through variables linked to their electromagnetic response. This physical connection underlies the three roles of the representation. Low-order descriptors summarize resonance trends within individual branches, while local complex coefficients retain the amplitude and phase information needed to learn resonance lineshapes. The same coordinates also regularize inverse reconstruction when used as the model output. Structural comparison, spectral prediction and inverse reconstruction thus share a common physical description.

This connection guides the model architecture. The resonance branch learns a mapping from low-order complex coefficients associated with harmonic coupling to the parameters of a pole-based lineshape consistent with temporal coupled-mode theory. Physics-informed neural networks typically incorporate governing equations into the loss, whereas Fourier neural operators use the Fourier basis to parameterize a learned operator. Here, the input representation and output lineshape constrain the learning process. The ablations support a joint improvement from the local complex Fourier input and the Fano-parameterized output. This design may also apply to other spectra dominated by isolated resonances, such as guided-mode and surface-lattice resonances, provided that the target resonance can be assigned to a known spectral window.

Whether the relationships learned in these shared coordinates extend to unseen structure families is a further test of the representation. All seven families appear in training in the main forward and inverse evaluations, so these results establish joint learning within a known design space. The LOFO results further suggest that the model learns transferable relationships between structural changes and spectral responses from the other six families. The substantial improvement in full-spectrum prediction after adding just one target-family sample supports this interpretation. Broader training coverage in the relevant Fourier-coefficient space would allow further assessment of how these learned relationships apply across structure families. Material differences can, in principle, also be encoded through the permittivity distribution, with dispersion and loss included where relevant, providing a basis for exploring spectral learning across material systems.

The comparisons here use a common lattice, material, film thickness and excitation, and broader generalization remains to be tested. The perturbative analysis assumes weak leakage and small perturbations, while the field tests validate linear trends for two branches. Inverse reconstruction also retains the non-uniqueness of the mapping from spectra to structures, so the model returns one plausible structure for a given spectrum.

The dielectric Fourier representation and resonance expert could be combined with physics-informed optimization strategies to improve spectral accuracy and design efficiency while preserving the shared structural coordinates. In parallel, we make publicly available a dataset of 2,015 simulated samples spanning seven dielectric metasurface families, together with their corresponding dielectric distributions and optical spectra. This dataset provides a reusable basis for investigating structure–spectrum relationships across different metasurface families, evaluating cross-family generalization, and exploring alternative data-driven models and inverse-design strategies. Together, these developments may facilitate the design of high-Q devices with tailored spectral responses for selective light–matter interactions, weak-signal enhancement and compact spectral processing, and motivate further exploration of applications in quantum communication, optical signal detection and integrated photonic chips.

## Methods

### Full-wave simulation and dataset construction

Transmission spectra of the seven silicon metasurface families were computed using three-dimensional finite-difference time-domain (FDTD) simulations in Lumerical. Each family uses a fixed unit-cell motif controlled by two geometric parameters, $P_1$ and $P_2$. These parameters were sampled on a nominal 17 × 17 grid. The lattice period (1,320 nm), silicon thickness (250 nm), substrate index (1.0), excitation, boundary conditions and monitors are common to all families; only the in-plane permittivity distribution changes.

We applied periodic boundary conditions in the x–y plane and perfectly matched layers along $z$. The structures were illuminated by a normally incident plane wave. Transmission was recorded at 2,001 wavelengths (Supplementary Table S2). The permittivity in the silicon layer, recorded by an index monitor, provides both the real-space input (resampled to 67 × 67) and the Fourier representation described below.

Several spectra contain more than one resonance or background dip. We defined a target wavelength window for each family from the parameter-resolved spectra and tracked one continuous qBIC branch within it. From this branch, we extracted the resonance position, minimum transmission, linewidth and $Q$. Of the 2,023 simulated structures, eight (seven CSR and one DAN) were excluded because the target resonance could not be identified, leaving 2,015.

The strict split separates source-grid geometry hashes and principal-scan parameter blocks, giving 1,556 training, 238 validation and 221 test samples. It is fixed for all models (Supplementary Note 1 and Supplementary Fig. S17).

### Low-order dielectric Fourier- descriptor analysis

The spatial mean was subtracted from each permittivity map before a two-dimensional FFT,

and the zero-frequency component was shifted to the centre to give coefficients $F_{mn}$ on the reciprocal-lattice grid. Details of the Fourier-grid construction and cropping procedure are provided in Supplementary Note 3.

ExtraTrees regressors trained on the 67 × 67 Fourier pixels identified candidate low-order regions for resonance position and ln $Q$. From these regions we defined $C_\Gamma$ (Eq. (6)) and $C_D$ (Eq. (7)), whose definitions were then frozen for all families, scan directions and slices.

Within each 17-point slice, we used ordinary least squares to fit linear relationships between each descriptor and resonance position or ln Q . One parameter was scanned while the other remained fixed. Reported $R^2$ values are averaged over the 17 slices of each family (Supplementary Note 3 and Supplementary Table S3).

**Physics-guided spectral learning**

Four configurations were built: Real bg, K bg, Real + K peak and K + K peak. The backbone is a multiscale convolutional encoder that processes the full input and its central 33 × 33 region and outputs 32 PCA coefficients of the spectrum. Its input is either the 1 × 67 × 67 real-space map or the 5 × 67 × 67 Fourier tensor of Eq. (13). The background-only models and the background stage of the expert models are trained against all 2,001 points in each family-specific target window. The PCA basis is fitted to these full training spectra.

The expert models add a convolutional branch that uses the central 5 × 9 × 9 Fourier block to predict $\lambda_0$, ln $Q$, $d$ and $q_f$. A differentiable Fano layer converts these parameters into a resonance term added to the backbone spectrum (Eq. (14)). Input statistics, target statistics and the PCA basis were fitted on the training set only.

Models were trained with AdamW, selected on the validation set and evaluated once on the test set; each configuration was run with five seeds. Hyperparameters and loss weights are listed in Supplementary Note 5.

We evaluated the models using full-window spectral MAE, resonance-position MAE, minimum-transmission MAE and the per-sample Pearson correlation between predicted and simulated spectra. We also computed local MAEs within ±8 nm and ±5 nm and a core MAE within ±0.5 nm of the labelled resonance. These local and core metrics were used for evaluation only (Supplementary Figs. S19–S21).

**Spectrum-driven Fourier-space inverse reconstruction**

The inverse model uses a one-dimensional convolutional network to encode the spectrum in the target window and the normalized wavelength. It predicts both family probabilities and the complex Fourier coefficients of the unit cell.

Family labels provide auxiliary supervision during training; at inference, only the predicted probabilities are used. The Fourier target is the orthonormal real FFT of the binary mask logits (±4), standardized with training-set statistics (Supplementary Note 7).

The predicted coefficients are inverse-transformed and passed through a sigmoid to give a 67 × 67 foreground-probability map. We compare Fourier-only supervision, joint Fourier and real-space supervision, and direct real-space decoding (Supplementary Figs. S26 and S27).

**Table 1 |** Summary of key metrics for the four models. Values are mean ± s.d. over five seeds.

| Model | Full-window spectral MAE | Local MAE (±8 nm) | Minimum-transmission MAE | Resonance-position MAE (nm) | ln Q MAE | Median Pearson correlation R |
|---|---|---|---|---|---|---|
| Real bg | 0.0329± 0.0011 | 0.0665± 0.0020 | 0.5870± 0.0341 | 2.9998± 0.0798 | 0.5415± 0.0426 | 0.2983± 0.0265 |
| K bg | 0.0375± 0.0063 | 0.0685± 0.0068 | 0.6617± 0.0490 | 3.2226± 0.3410 | 0.6728± 0.2260 | 0.2624± 0.0685 |
| Real + K peak | 0.0219± 0.0006 | 0.0448± 0.0023 | 0.0535± 0.0031 | 0.9981± 0.0594 | 0.2886± 0.0104 | 0.7377± 0.0622 |
| K + K peak | 0.0194± 0.0007 | 0.0454± 0.0022 | 0.0471± 0.0051 | 0.9473± 0.0653 | 0.2954± 0.0174 | 0.7148± 0.0322 |

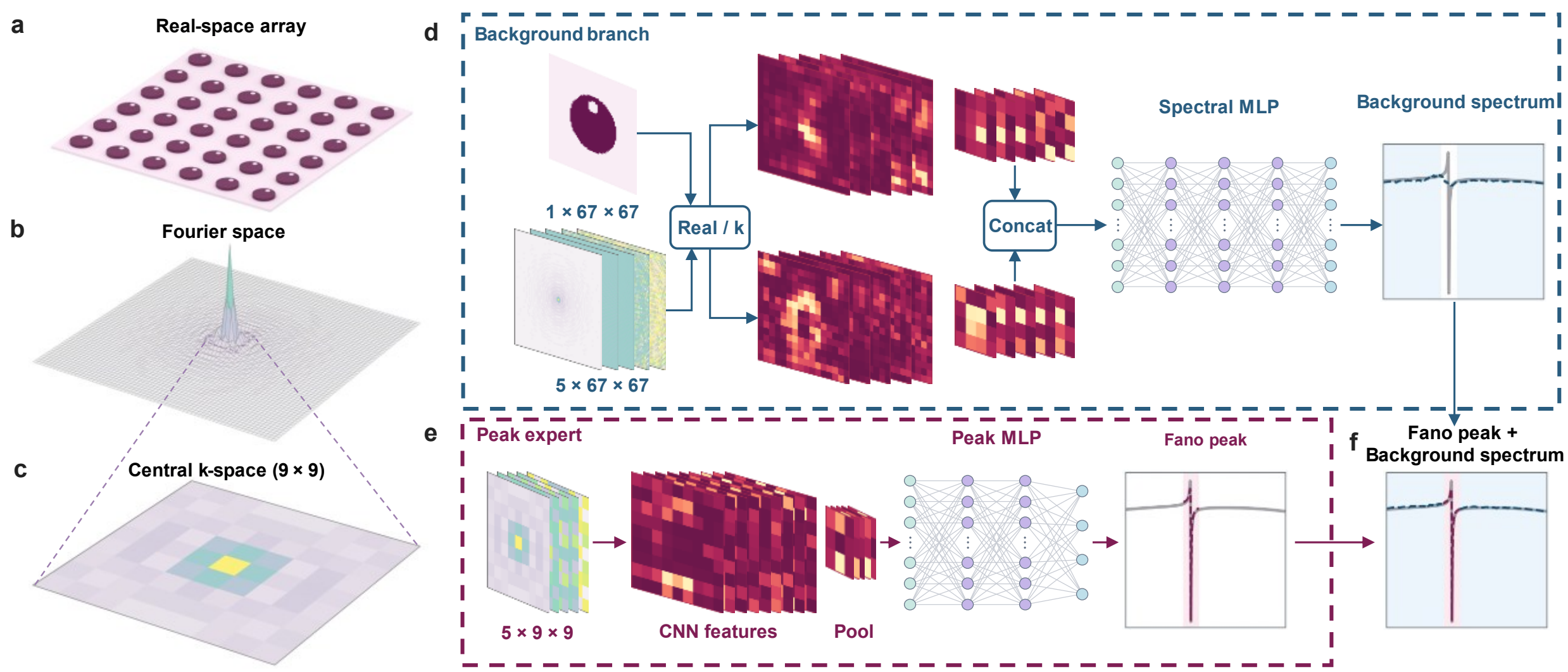

**Fig. 1 | Dielectric Fourier-space framework for qBIC spectral learning.** a, Schematic of the periodic metasurface array in real space. b, Fourier-space representation of the structure. The central region corresponds to low spatial frequencies. c, The 9 × 9 region extracted from the centre of Fourier space. Its five channels serve as input to the peak expert. d, Background branch with either a real-space input (1 × 67 × 67) or a Fourier-space input (5 × 67 × 67). Pooled features from the full input and its central 33 × 33 region are concatenated. A spectral multilayer perceptron (MLP) predicts 32 principal component analysis (PCA) coefficients to reconstruct the background. Grey solid and blue dashed curves show the raw FDTD spectrum and background prediction, respectively. e, Peak expert. The central Fourier-space data (5 × 9 × 9) pass through a convolutional neural network (CNN) and a pooling layer. A peak MLP predicts the resonance centre, ln Q, depth and Fano asymmetry parameter for a differentiable Fano layer. The pink dashed curve illustrates the predicted resonance profile. f, Final spectrum reconstruction. The PCA background and Fano resonance components are combined to reconstruct the spectrum. Blue and pink dashed segments mark the broadband and resonance regions of the combined prediction.

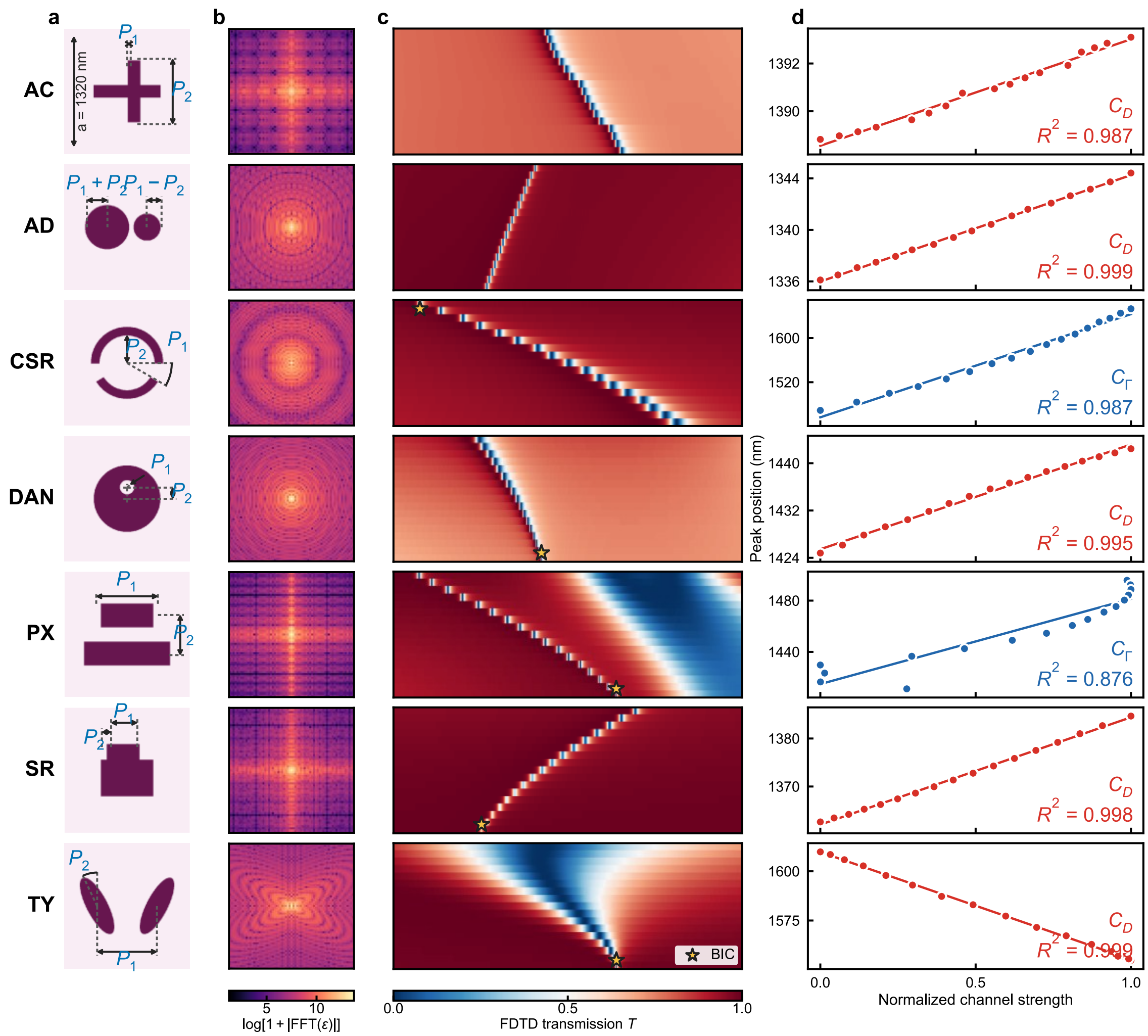

**Fig. 2 | Seven qBIC metasurface families on a shared Fourier grid.** a, Unit cells and geometric parameters $P_1$ and $P_2$ (lattice period $a$ = 1,320 nm). b, Log-amplitude of the dielectric Fourier coefficients of representative structures. c, Raw FDTD transmission spectra along one parameter slice of each family (horizontal axis, wavelength; vertical axis, scanned parameter); stars mark the symmetry-protected BIC. d, Resonance position versus the normalized descriptor, $C_\Gamma$ for CSR and PX and $C_D$ for the other families. Each slice contains 17 points; $R^2$ is given for the slice shown.

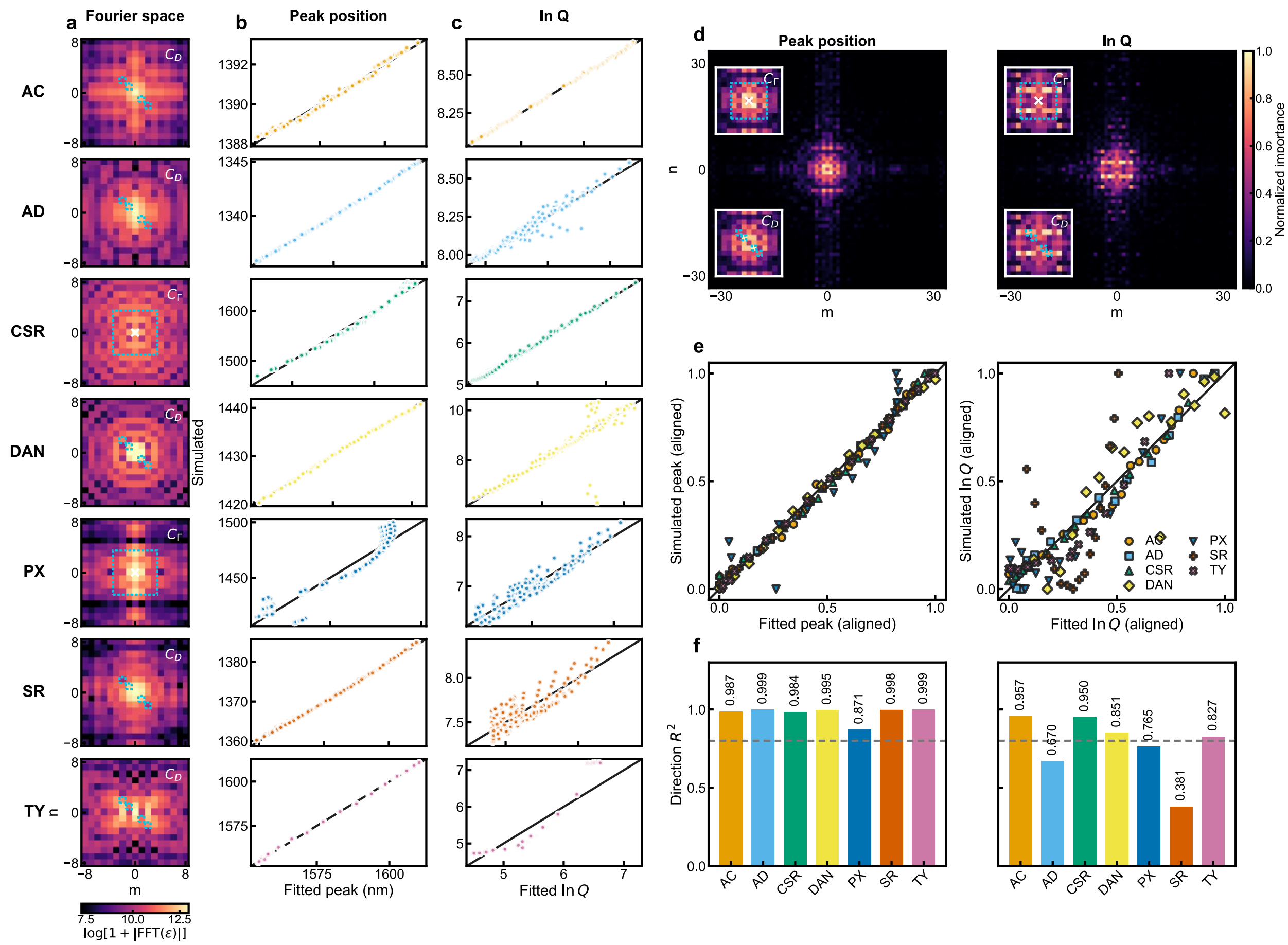

**Fig. 3 | Low-order Fourier descriptors of the qBIC resonance.** a, Dielectric Fourier maps of the seven families with the descriptor region outlined. b,c, Fitted versus simulated resonance position (b) and ln $Q$ (c) for all samples of each family. d, ExtraTrees importance maps for resonance position (left) and ln $Q$ (right); insets mark the $C_\Gamma$ and $C_D$ regions. e, All families after per-slice affine alignment of fitted and simulated values. f, Mean within-branch $R^2$ for resonance position (left) and ln $Q$ (right); dashed lines mark $R^2 = 0.8$. All fits are within-branch and in-sample.

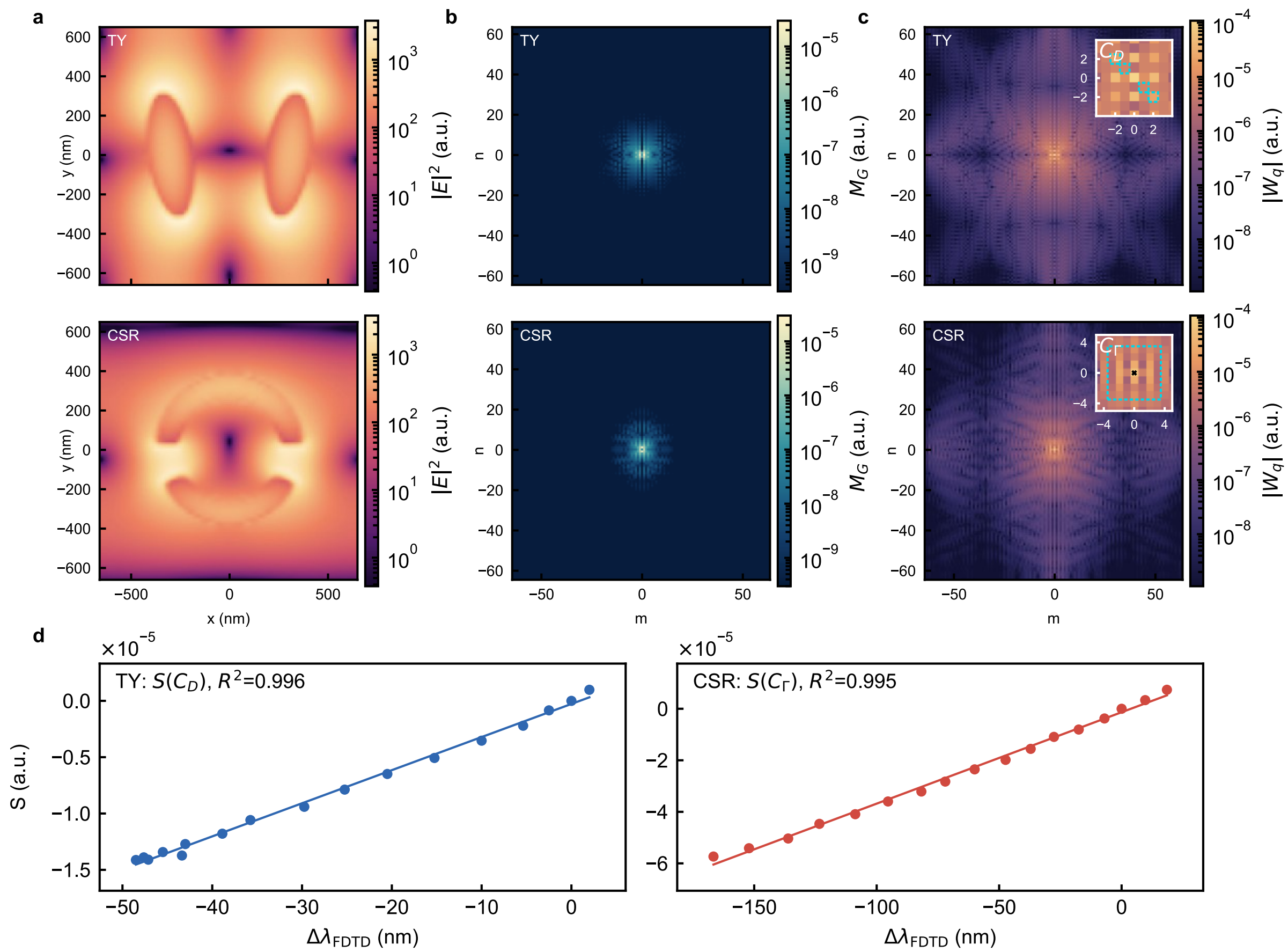

**Fig. 4 | Field-level test of the Maxwell–Fourier picture for the TY and CSR branches.** a, Resonant $|\mathbf{E}|^2$ at the mid-plane of the silicon layer. b, Modal Fourier weight $M_\mathbf{G}$ integrated over $z$. c, Magnitude $|W_\mathbf{q}|$ of the Fourier-harmonic overlap for the reference mode; insets mark the four $C_\mathrm{D}$ wavevectors (TY) and the non-DC central 7 × 7 block of $C_\Gamma$ (CSR). d, Perturbative response $S_i$ versus the resonance shift Δλ along each 17-state branch ($R^2$ = 0.996 for TY and 0.995 for CSR).

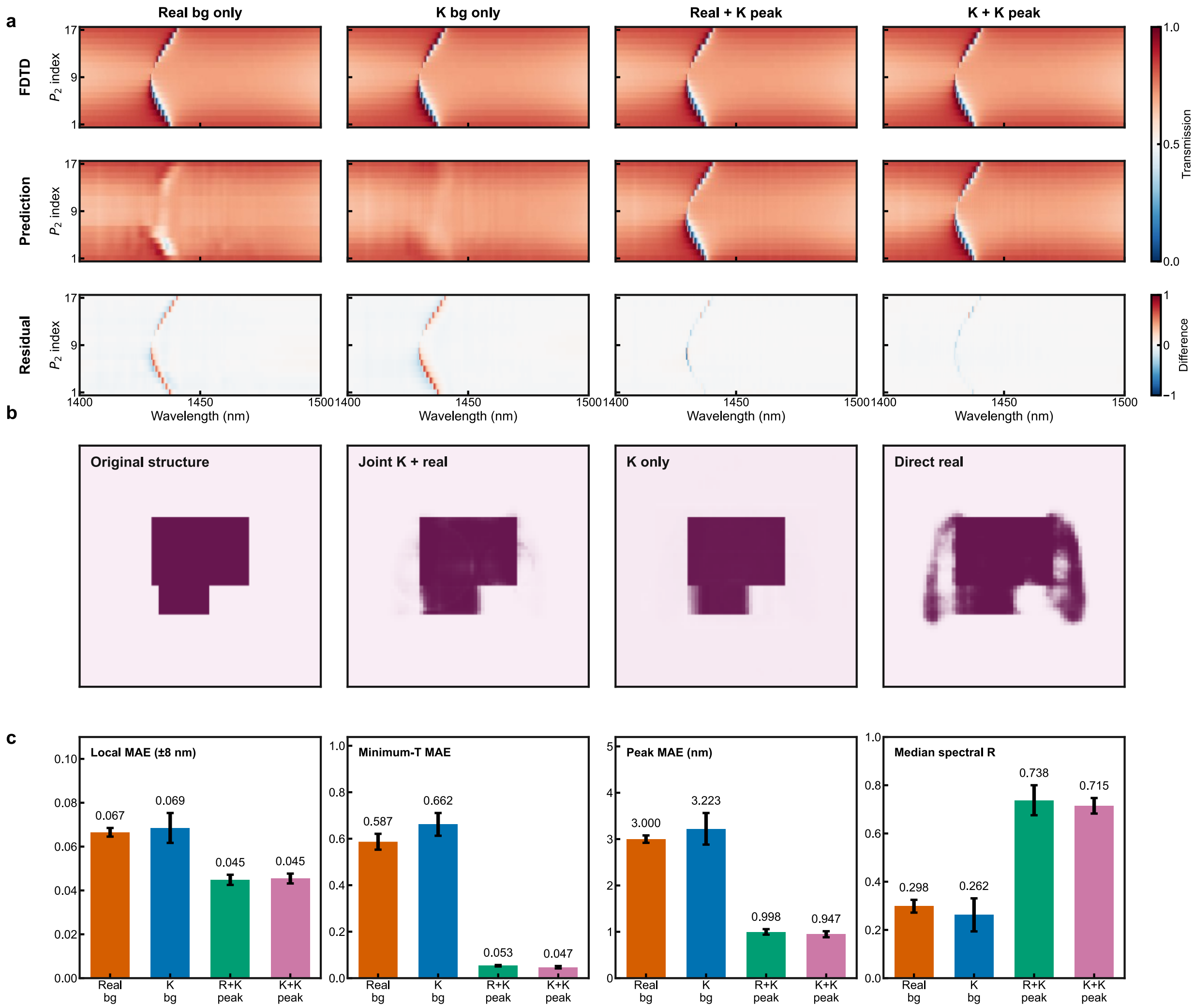

**Fig. 5 | Forward prediction with the resonance expert and inverse reconstruction.** a, Raw FDTD transmission spectra, predictions of the four models and residuals (prediction − FDTD) along a DAN slice ($P_1$ index 10; $P_2$ index 1–17). b, Test structure S05_SR_008 and its single-seed reconstructions by joint Fourier and real-space supervision, Fourier-only supervision and direct real-space decoding. c, Local MAE (±8 nm), minimum-transmission MAE, resonance-position MAE and median Pearson correlation of the four models on the 221 test samples (mean ± s.d. over five seeds).

## References

1. Peurifoy, J. et al. Nanophotonic particle simulation and inverse design using artificial neural networks. *Sci. Adv.* **4**, eaar4206 (2018).

2. So, S. & Rho, J. Designing nanophotonic structures using conditional deep convolutional generative adversarial networks. *Nanophotonics* **8**, 1255–1261 (2019).

3. Ma, W. et al. Deep learning for the design of photonic structures. *Nat. Photonics* **15**, 77–90 (2021).

4. Malkiel, I. et al. Plasmonic nanostructure design and characterization via Deep Learning. *Light Sci. Appl.* **7**, 60 (2018).

5. Sajedian, I., Kim, J. & Rho, J. Finding the optical properties of plasmonic structures by image processing using a combination of convolutional neural networks and recurrent neural networks. *Microsyst. Nanoeng.* **5**, 27 (2019).

6. Wei, Z. et al. Ultraprecision, high-capacity, and wide-gamut structural colors enabled by a mixture probability sampling network. *Light Sci. Appl.* **15**, 164 (2026).

7. Rahaman, N. et al. On the spectral bias of neural networks. In *Proc. 36th International Conference on Machine Learning* Vol. 97 (eds Chaudhuri, K. & Salakhutdinov, R.) 5301–5310 (PMLR, 2019).

8. Hsu, C. W., Zhen, B., Stone, A. D., Joannopoulos, J. D. & Soljačić, M. Bound states in the continuum. *Nat. Rev. Mater.* **1**, 16048 (2016).

9. Hsu, C. W. et al. Observation of trapped light within the radiation continuum. *Nature* **499**, 188–191 (2013).

10. Koshelev, K., Lepeshov, S., Liu, M., Bogdanov, A. & Kivshar, Y. Asymmetric metasurfaces with high-*Q* resonances governed by bound states in the continuum. *Phys. Rev. Lett.* **121**, 193903 (2018).

11. Limonov, M. F., Rybin, M. V., Poddubny, A. N. & Kivshar, Y. S. Fano resonances in photonics. *Nat. Photonics* **11**, 543–554 (2017).

12. Kodigala, A. et al. Lasing action from photonic bound states in continuum. *Nature* **541**, 196–199 (2017).

13. Tittl, A. et al. Imaging-based molecular barcoding with pixelated dielectric metasurfaces. *Science* **360**, 1105–1109 (2018).

14. Yesilkoy, F. et al. Ultrasensitive hyperspectral imaging and biodetection enabled by dielectric metasurfaces. *Nat. Photonics* **13**, 390–396 (2019).

15. Liu, Z. et al. High-*Q* quasibound states in the continuum for nonlinear metasurfaces. *Phys. Rev. Lett.* **123**, 253901 (2019).

16. Chen, Y. et al. Observation of intrinsic chiral bound states in the continuum. *Nature* **613**, 474–478 (2023).

17. Zhen, B., Hsu, C. W., Lu, L., Stone, A. D. & Soljačić, M. Topological nature of optical bound states in the continuum. *Phys. Rev. Lett.* **113**, 257401 (2014).

18. Qin, H. et al. Arbitrarily polarized bound states in the continuum with twisted photonic crystal slabs. *Light Sci. Appl.* **12**, 66 (2023).

19. Ma, X. et al. Strategical deep learning for photonic bound states in the continuum. *Laser Photonics Rev.* **16**, 2100658 (2022).

20. Wang, L., Wang, W., Dong, Q., Wang, L. & Gao, L. Deep learning enabled inverse design of bound states in the continuum with ultrahigh Q factor. *J. Opt. Soc. Am. B* **41**, A146–A151 (2024).

21. Joannopoulos, J. D., Johnson, S. G., Winn, J. N. & Meade, R. D. *Photonic Crystals: Molding the Flow of Light* 2nd edn (Princeton Univ. Press, 2008).

22. Fan, S. & Joannopoulos, J. D. Analysis of guided resonances in photonic crystal slabs. *Phys. Rev. B* **65**, 235112 (2002).

23. Gao, X. et al. Formation mechanism of guided resonances and bound states in the continuum in photonic crystal slabs. *Sci. Rep.* **6**, 31908 (2016).

24. Lee, S.-G. & Magnusson, R. Band flips and bound-state transitions in leaky-mode photonic lattices. *Phys. Rev. B* **99**, 045304 (2019).

25. Lee, S.-G., Kim, S.-H. & Kee, C.-S. Metasurfaces with bound states in the continuum enabled by eliminating first Fourier harmonic component in lattice parameters. *Phys. Rev. Lett.* **126**, 013601 (2021).

26. Lee, S.-G., Kim, S.-H. & Kee, C.-S. Fourier-component engineering to control light diffraction beyond subwavelength limit. *Nanophotonics* **10**, 3917–3925 (2021).

27. Glauser, Y. M. et al. Fourier pixels for bidirectional light control. *Nature* **655**, 87–92 (2026).

28. Li, Z. et al. Fourier neural operator for parametric partial differential equations. In *Proc. 9th International Conference on Learning Representations* (ICLR, 2021).

29. Raissi, M., Perdikaris, P. & Karniadakis, G. E. Physics-informed neural networks: a deep learning framework for solving forward and inverse problems involving nonlinear partial differential equations. *J. Comput. Phys.* **378**, 686–707 (2019).

30. Karniadakis, G. E. et al. Physics-informed machine learning. *Nat. Rev. Phys.* **3**, 422–440 (2021).

31. Wang, J. T., You, J. W. & Panoiu, N. C. Giant second-harmonic generation in monolayer $MoS_2$ boosted by dual bound states in the continuum. *Nanophotonics* **13**, 3437–3448 (2024).

32. van Loon, T. et al. Refractive index sensing using quasi-bound states in the continuum in silicon metasurfaces. *Opt. Express* **32**, 14289–14299 (2024).

33. Huang, Z. et al. Quasi-bound state in the continuum in a dielectric double-gap split-ring metasurface structure with large split angles. *Opt. Mater. Express* **14**, 1484–1498 (2024).

34. Tuz, V. R. et al. High-quality trapped modes in all-dielectric metamaterials. *Opt. Express* **26**, 2905–2916 (2018).

35. Campione, S. et al. Broken symmetry dielectric resonators for high quality factor Fano metasurfaces. *ACS Photonics* **3**, 2362–2367 (2016).

36. Gölz, T. et al. Revealing mode formation in quasi-bound states in the continuum metasurfaces via near-field optical microscopy. *Adv. Mater.* **36**, 2405978 (2024).

37. Ashcroft, N. W. & Mermin, N. D. *Solid State Physics* (Holt, Rinehart and Winston, 1976).

38. Johnson, S. G. et al. Perturbation theory for Maxwell's equations with shifting material boundaries. *Phys. Rev. E* **65**, 066611 (2002).

**Code availability**

Code for Fourier descriptor analysis, forward spectral prediction, inverse geometry reconstruction and model evaluation is available at https://github.com/leishuangteng/FFTBIC-Core.

**Data availability**

The unit-cell permittivity maps, Fourier input tensors, raw transmission spectra in the target windows, dataset splits and model evaluation results used in this study are available at https://github.com/leishuangteng/FFTBIC-Core.

**Acknowledgements**

This work was financially supported by Chinese Academy of Sciences (Grant Nos. XDB0580000.), National Natural Science Foundation of China (Grant Nos. 12393833, 12227901, U2241219, 12174416, 11991063.), and the Science and Technology Commission of Shanghai Municipality (Grant No. 23JC1404100). The simulations and model training were performed on the robotic AI-Scientist platform of Chinese Academy of Sciences.

**Author contributions**

L.Y., T.L., and W.L. conceived the project and designed experiments. S.L., and L.Y. performed the simulation data acquisition and model training. L.Y., and W.L. conducted the theoretical analysis. S.L., L.Y., T.L. and W.L. analysed the results. All authors contributed to interpretation of the results and the writing of the manuscript.

**Competing interests**

The authors declare no competing interests.

## Supplementary Information

## Learning qBIC Resonances across Metasurface Families in Dielectric Fourier Space

This file contains Supplementary Notes 1–7, Supplementary Figs. S1–S27 and Supplementary Tables S1–S3. Unless stated otherwise, all analyses use the 2,015 valid samples and the fixed strict split of 1,556 training, 238 validation and 221 test samples.

### Supplementary Note 1 | Dataset, simulation settings and continuous spectral branches

The seven families (AC, AD, CSR, DAN, PX, SR and TY) were simulated on nominal 17 × 17 grids of the geometric parameters $P_1$ and $P_2$. Eight structures were excluded because the target resonance could not be identified (seven CSR samples and DAN231), leaving 2,015 valid samples. Table S1 lists the parameter definitions, ranges, target windows and split counts.

**Supplementary Table S1 |** Geometric parameters, target windows and strict-split composition of the seven families.

| Structure | $P_1$ | $P_2$ | $P_1$ range | $P_2$ range | BIC window (nm) | Scan parameter | Total | Train | Val | Test |
|---|---|---|---|---|---|---|---|---|---|---|
| AC | Δx | ΔL | 50–100 nm | 0–100 nm | 1380–1400 | $P_2$ | 289 | 238 | 17 | 34 |
| AD | R0 | ΔR | 180–194 nm | 30–60 nm | 1320–1380 | $P_1$ | 289 | 221 | 34 | 34 |
| CSR | Δθ | Rin | 3–30 deg | 256–320 nm | 1450–1700 | $P_2$ | 282 | 214 | 34 | 34 |
| DAN | r_hole | Δx | 50–100 nm | 24–216 nm | 1400–1500 | $P_1$ | 288 | 220 | 34 | 34 |
| PX | d | Δp | 360–440 nm | 200–500 nm | 1400–1550 | $P_2$ | 289 | 204 | 51 | 34 |
| SR | Δa | Δb | 262.5–425 nm | 0–125 nm | 1350–1400 | $P_1$ | 289 | 221 | 34 | 34 |
| TY | d | θ | 600–660 nm | 5–50 deg | 1450–1700 | $P_2$ | 289 | 238 | 34 | 17 |

Table S2 lists the common FDTD settings. $P_2$ is scanned for AC, CSR, PX and TY, and $P_1$ for AD, DAN and SR; the other parameter is fixed within each branch. The same conventions apply to main-text Fig. 2.

| **Supplementary Table S2 \|** Three-dimensional FDTD settings. | |
|---|---|
| **Item** | **Setting** |
| Simulation model | Three-dimensional Lumerical FDTD; identical modelling framework for all seven families. |
| Simulation domain and boundaries | x,y ∈ [−0.66, 0.66] μm; z ∈ [−10, 7] μm; periodic boundaries along x/y; eight-layer PML along z. |
| Materials and thickness | Si (Silicon)–Palik; metasurface thickness, 250 nm; substrate refractive index, n = 1.0; background refractive index, n = 1.0. |
| Source | Normally incident periodic plane wave; wavelength range, 1,300–1,700 nm; polarization angle, 0°; source plane, z = 3 μm; Backward (along −z); θ = φ = 0°. |
| Mesh | Global auto non-uniform mesh; mesh accuracy = 2; local mesh, dx = dy = dz = 10 nm; local range, x,y ∈ [−0.60, 0.60] μm and z ∈ [−0.10, 0.30] μm. |
| Time and termination | Simulation time, 50 ps; auto-shutoff minimum = $1 \times 10^{-7}$; stability factor = 0.99. |
| Monitors | Reflection monitor R, z = +4 μm; transmission monitor T, z = −4 μm; dielectric/refractive-index monitor, z = 100 nm. |
| Spectral sampling | Uniform sampling of 2,001 points over the family-specific windows; the complete 2,001-point array is retained for every input spectrum. |

The learning targets are 2,001-point spectra within the family-specific windows of Table S1; these windows differ from the 1,300–1,700 nm acquisition range of the monitor. Resonance labels follow the selected continuous branch within each window, and structures without a reliable label were excluded rather than assigned to another branch. Figures S1–S7 show the branches for several fixed values of the complementary parameter and extend main-text Fig. 2c.

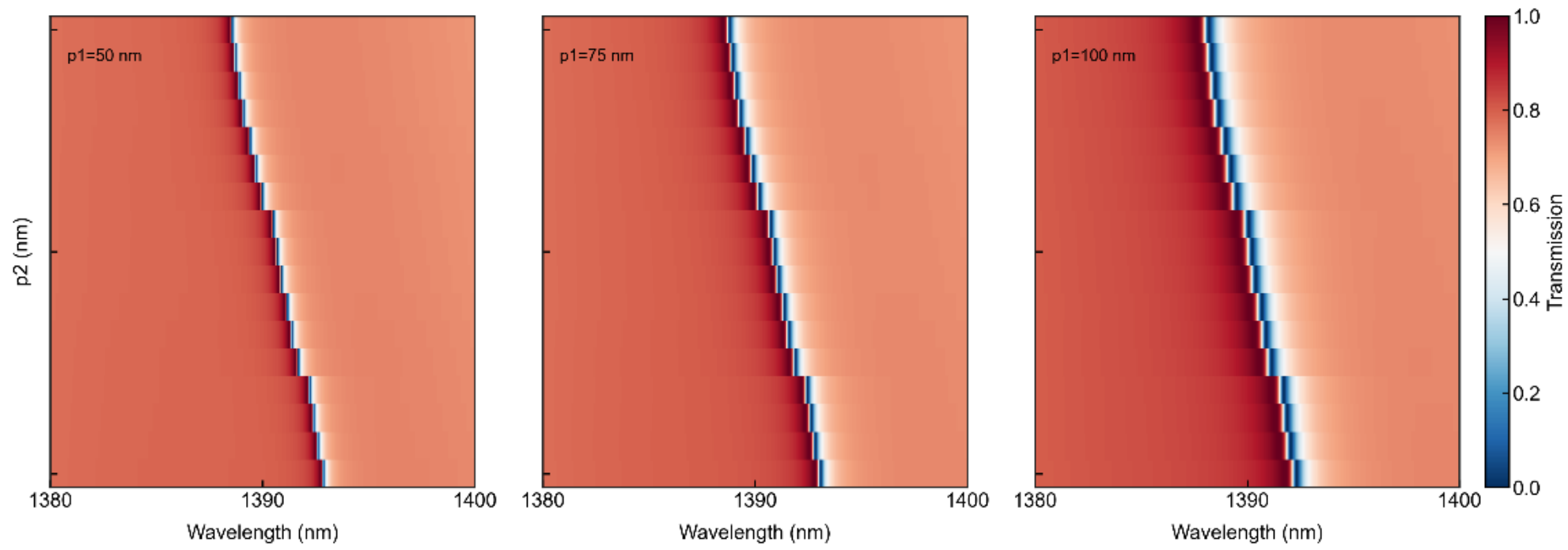

**Supplementary Fig. S1 | AC transmission spectra along the $P_2$ scan. Columns correspond to different fixed values of $P_1$.**

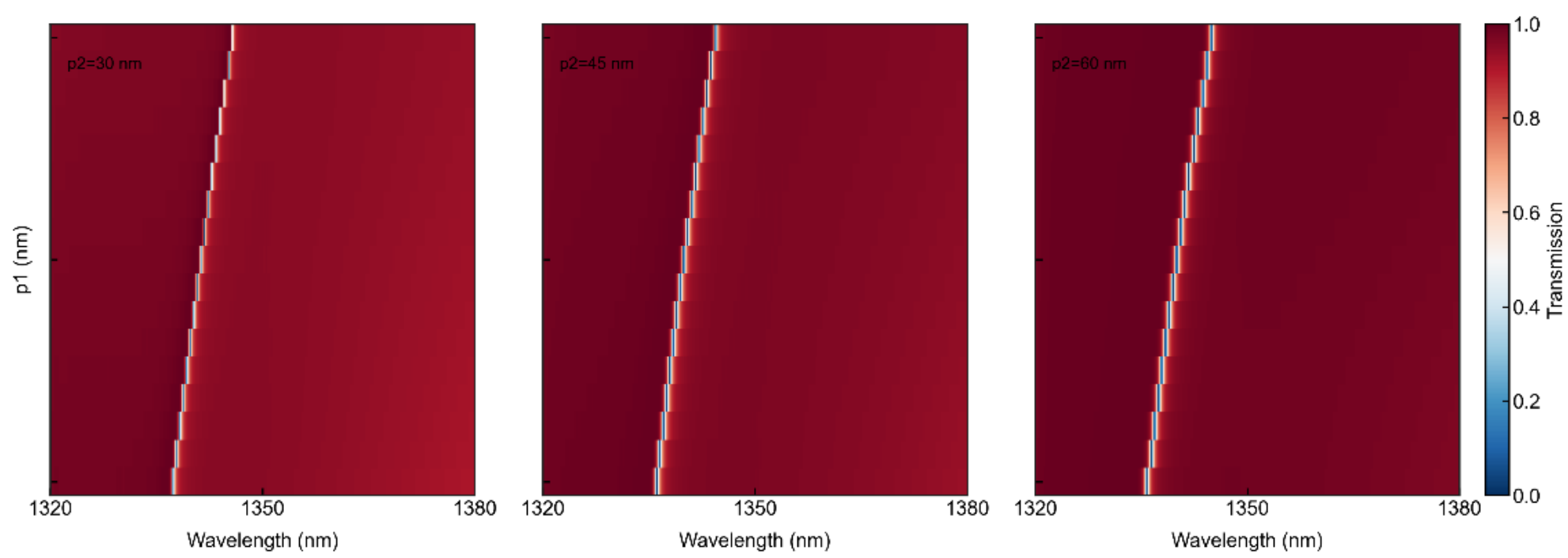

**Supplementary Fig. S2 | AD transmission spectra along the $P_1$ scan. Columns correspond to different fixed values of $P_2$.**

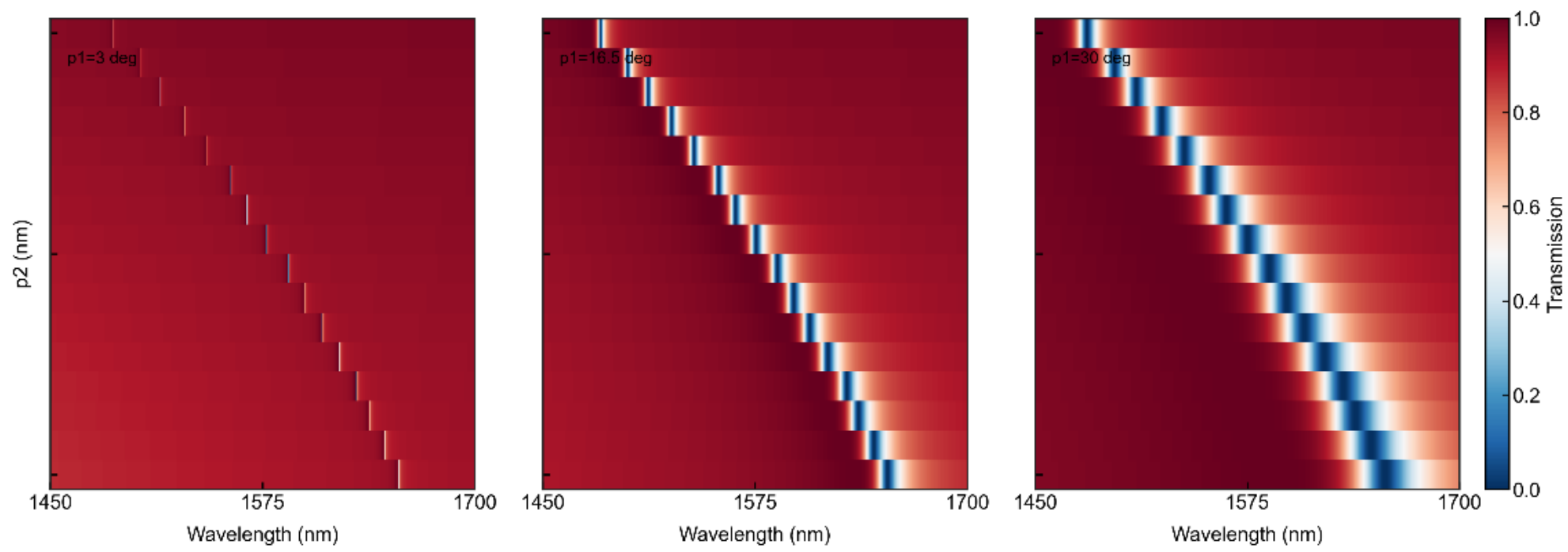

**Supplementary Fig. S3 | CSR transmission spectra along the $P_2$ scan. Columns correspond to different fixed values of $P_1$.**

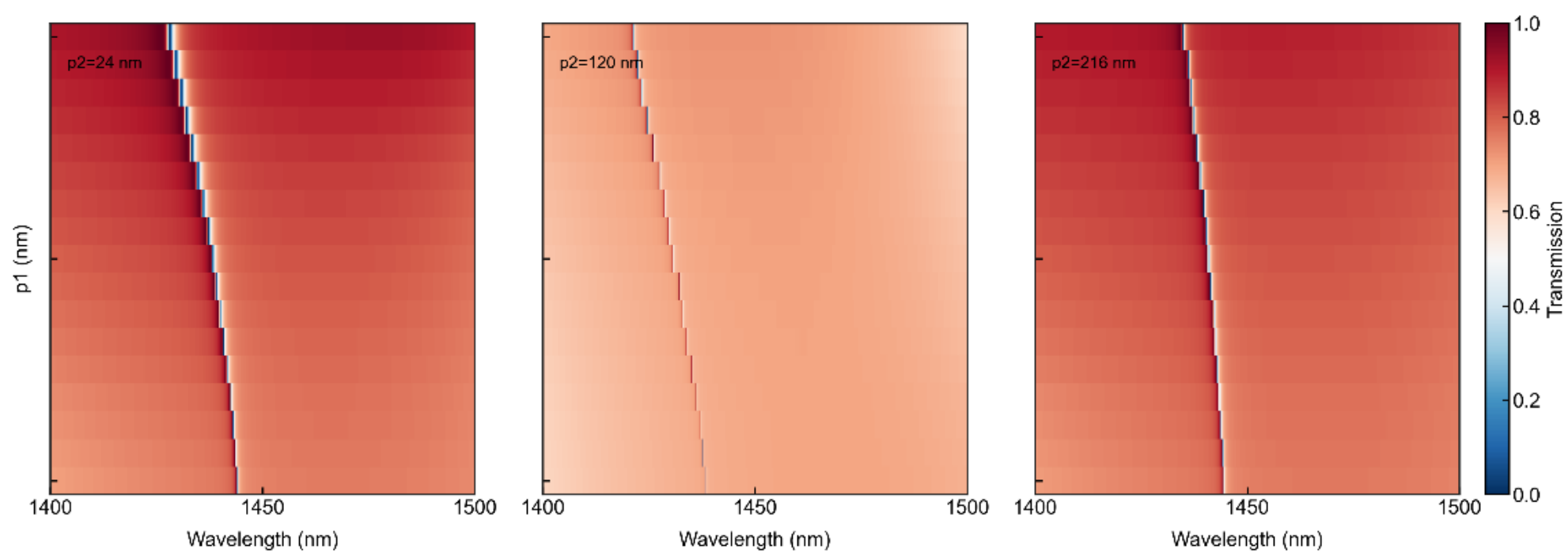

**Supplementary Fig. S4 | DAN transmission spectra along the $P_1$ scan. Columns correspond to different fixed values of $P_2$.**

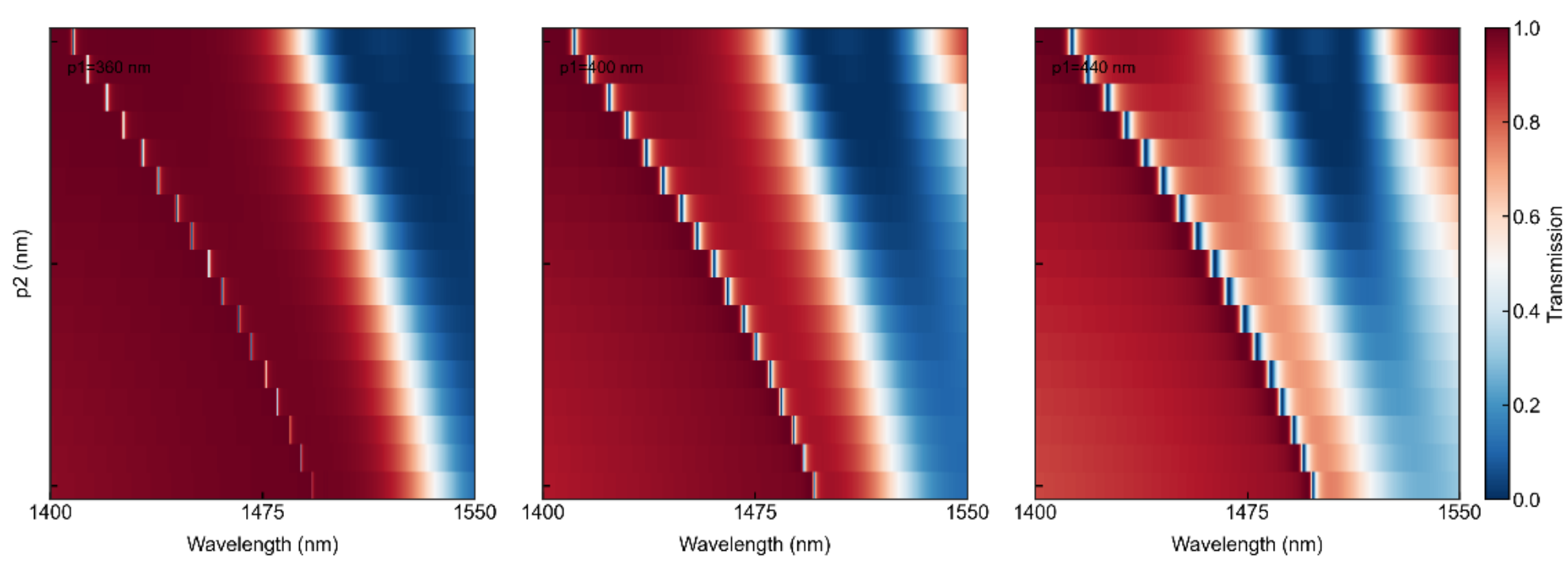

**Supplementary Fig. S5 | PX transmission spectra along the $P_2$ scan. Columns correspond to different fixed values of $P_1$.**

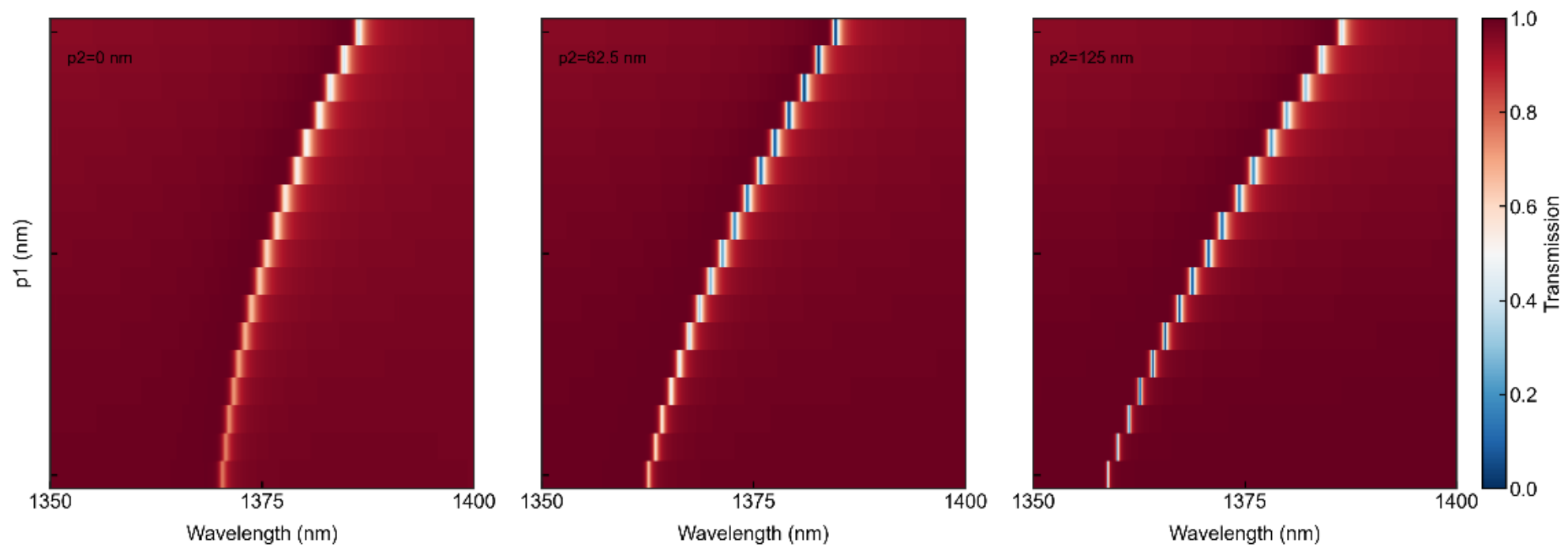

**Supplementary Fig. S6 | SR transmission spectra along the $P_1$ scan. Columns correspond to different fixed values of $P_2$.**

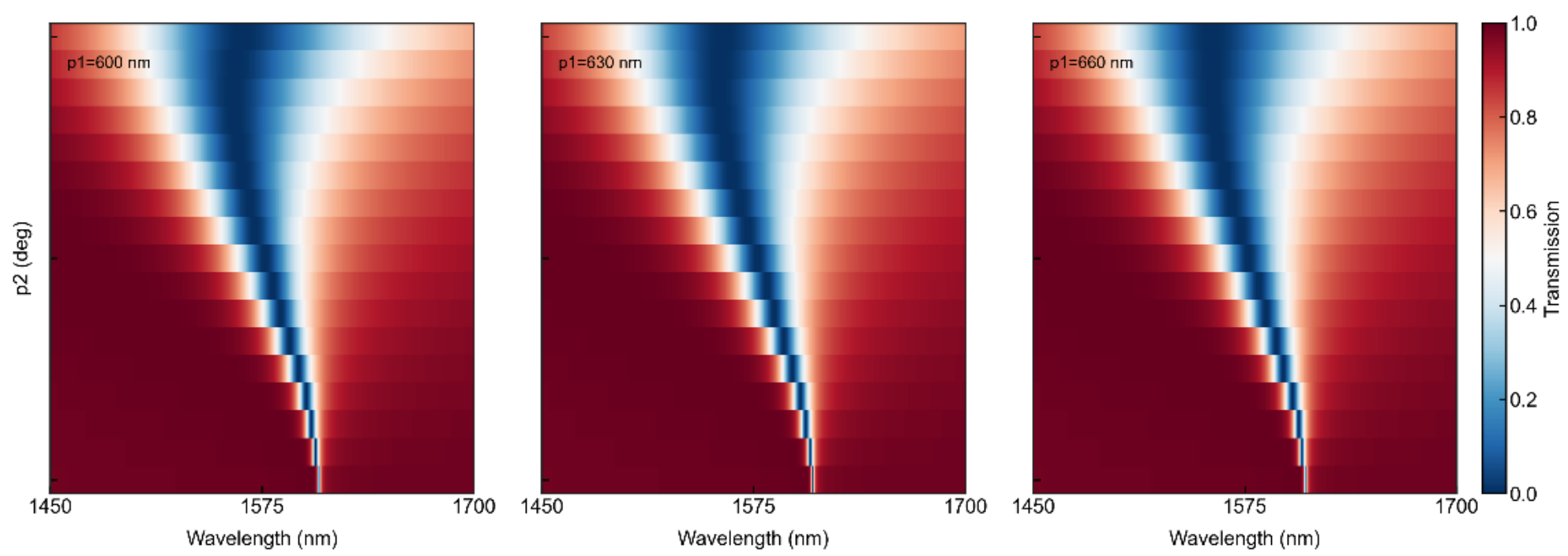

**Supplementary Fig. S7 | TY transmission spectra along the $P_2$ scan. Columns correspond to different fixed values of $P_1$.**

**Supplementary Note 2 | Compact Maxwell–Fourier derivation**

For a unit cell of area $A_{cell}$, the in-plane dielectric expansion and its coefficients are

$$\varepsilon(\rho, z) = \sum_q \varepsilon_q(z) e^{(iq\cdot\rho)}, \qquad (S1)$$

$$\varepsilon_q(z) = 1/A_{cell} \int_{A_{cell}} \varepsilon(\rho, z) e^{(-iq\cdot\rho)}\, d^2\rho, \qquad (S2)$$

Here $q = mb_1 + nb_2$ is a reciprocal-lattice vector. Expanding the Bloch field in harmonics G gives the convolutional Maxwell equation below; $L_G$ includes the z-dependent differential operator for a slab. Its dielectric coupling matrix is $B_{G,G\prime} = \varepsilon_{G-G\prime}$.

$$L_G E_G = (\omega^2/c^2) \sum_{G'} \varepsilon_{(G-G')}\, E_{(G')}, \qquad (S3)$$

Write $A\, E_R \;=\; \tilde{\lambda}\, B\, E_R$, with $\tilde{\lambda} \;=\; \tilde{\omega}^2/c^2$. At fixed A, first-order variation about the reference state r and projection onto the left eigenfield $E_L$ give

$$\Delta\tilde{\omega}_i = -(\tilde{\omega}_r/2)\big(E_L^\dagger \Delta B_i E_R\big)/\big(E_L^\dagger B_r E_R\big), \qquad (S4)$$

The numerator can be grouped by wavevector transfer q. The common unit-cell area factor cancels against the normalization when the same Fourier convention is used in both:

$$E_L^\dagger \Delta B_i E_R = \sum_q \int dz\, \Delta\varepsilon_{(i,q)}(z) \sum_G E^*_{(L,G)}(z) \cdot E_{(R,G-q)}(z), \qquad (S5)$$

For a perturbation uniform through the patterned thickness, the z integral defines $W_q = \Sigma_G \int E^*_{r,G} \cdot E_{r,G-q}$ dz over that region. Under weak leakage, $E_L \approx E_R$ and $T \;=\; \int\; E_r^* \;\cdot\; \varepsilon_r\, E_r\, dV$, giving the frequency and wavelength shifts

$$\Delta\omega_{(r,i)} \approx -\big(\omega_0/(2T)\big) Re\left[\sum_q \Delta\varepsilon_{(i,q)}\, W_q^{(r)}\right], \qquad (S6)$$

$$\Delta\lambda_i \approx \big(\lambda_0/(2T)\big) S_i, \qquad (S7)$$

Here $S_i = Re[\Sigma_q \Delta\varepsilon_{i,q} W_q]$. If Δε depends on z, it must remain inside the integral in Eq. (S5). Strong leakage, dispersion, moving discontinuous boundaries or modal hybridization require the corresponding generalized normalization, boundary perturbation or multimode treatment; the field tests below establish branchwise trend agreement under the stated approximation. Low-order modal localization supports substantial overlap at small q, but does not imply monotonic decay of $|W_q|$ with $|q|$.

For structure-side descriptors, $F_{mn}$ is the centered FFT of ε minus its spatial mean. The two fixed compressions exclude DC:

$$C_\Gamma = 1/48 \sum_{-3\le m,n\le 3;(m,n)\ne(0,0)} ln(1 + |F_{mn}|)\,, \tag{S8}$$

$$C_D = \sum_{(m,n)\in D} |F_{mn}|^2\,, D = (1,-1),(-1,1),(2,-2),(-2,2), \tag{S9}$$

These descriptors omit the modal weights and cross-component phases. For $\tilde{\omega} = \omega_r - i\gamma$, radiative Q instead depends on outgoing-channel amplitudes:

$$Q = \omega_r/\kappa = \omega_r/(2\gamma), \tag{S10}$$

$$d_\mu = d_{(\mu,0)} + \sum_q R_{(\mu q)}\,\Delta\varepsilon_q + O(\Delta\varepsilon^2), \tag{S11}$$

$$Q_{(rad)} \approx \omega_r/\left(2\sum_\mu \left|d_{(\mu,0)} + \sum_q R_{(\mu q)}\,\Delta\varepsilon_q\right|^2\right), \tag{S12}$$

The index μ labels a normalized outgoing diffraction, polarization and propagation channel, and $R_{\mu q}$ is its complex perturbative overlap. Coherent summation introduces the cross terms

$$\left|\sum_q a_q\right|^2 = \sum_q \left|a_q\right|^2 + \sum_{q\ne q'} a_q a^*_{(q')}\,, a_q = R_{(\mu q)}\Delta\varepsilon_q, \tag{S13}$$

Thus $ln\ Q_{rad} \approx C - ln\ \Sigma_{\mu} |d_{(\mu,0)} + \Sigma_q R_{(\mu q)} \Delta\varepsilon_q|^2$, corresponding to main-text Eq. (8). Amplitude-only $C_{\Gamma}$ and $C_D$ cannot recover this interference universally. The radiative expression assumes negligible absorption and other losses; spectral linewidth labels are not direct complex-eigenfrequency measurements. The central 9 × 9 complex input retains relative phases but does not explicitly supply $R_{\mu q}$.

**Supplementary Note 3 | Fixed descriptor definitions and branchwise validation**

The scalar descriptors are computed from the centered, mean-subtracted 529 × 529 dielectric FFT. The forward network instead receives the central 67 × 67 Fourier crop, with a central 9 × 9 expert crop. Real-space network inputs are 67 × 67 dielectric maps. Cropping reciprocal coefficients and downsampling a real-space image are different operations.

For resonance position, we compare the fixed $C_\Gamma$ and $C_D$ definitions along the designated scan direction for each family. Descriptors are ranked by the median within-branch $R^2$, then the mean $R^2$, minimum $R^2$ and normalized weighted MAE. Ordinary least squares is fitted separately within each branch. These are descriptive in-dataset descriptor fits, not held-out spectral-network predictions. Table S3 records the peak assignment and the separate ln Q descriptor used in main-text Fig. 3; the latter does not imply a universal law for radiative loss.

**Supplementary Table S3 |** Per-family Fourier descriptor assignments and branch statistics for resonance position and ln Q.

| Family | Scan / fixed | N | Peak descriptor | Peak mean $R^2$ | ln Q descriptor | ln Q mean $R^2$ |
|---|---|---|---|---|---|---|
| AC | P2 / P1 | 289 | $C_D$ | 0.987 | $C_D$ | 0.957 |
| AD | P1 / P2 | 289 | $C_D$ | 0.999 | $C_\Gamma$ | 0.670 |
| CSR | P2 / P1 | 282 | $C_\Gamma$ | 0.984 | $C_\Gamma$ | 0.950 |
| DAN | P1 / P2 | 288 | $C_D$ | 0.995 | $C_D$ | 0.851 |
| PX | P2 / P1 | 289 | $C_\Gamma$ | 0.871 | $C_D$ | 0.765 |
| SR | P1 / P2 | 289 | $C_D$ | 0.998 | $C_D$ | 0.381 |
| TY | P2 / P1 | 289 | $C_D$ | 0.999 | $C_\Gamma$ | 0.827 |

$R^2$ in Table S3 is the arithmetic mean of the 17 separate branch $R^2$ values in the selected direction. It is neither the median branch $R^2$ used for selection nor $R^2$ from one regression pooling raw points. The peak-position and ln Q descriptors are selected and reported separately. Figures S8–S14 show the complete peak-position branch fits; the representative-slice and aligned comparisons are shown in main-text Figs. 2d and 3e. The lower PX peak score and the nonuniform ln Q scores indicate the limits of the scalar description.

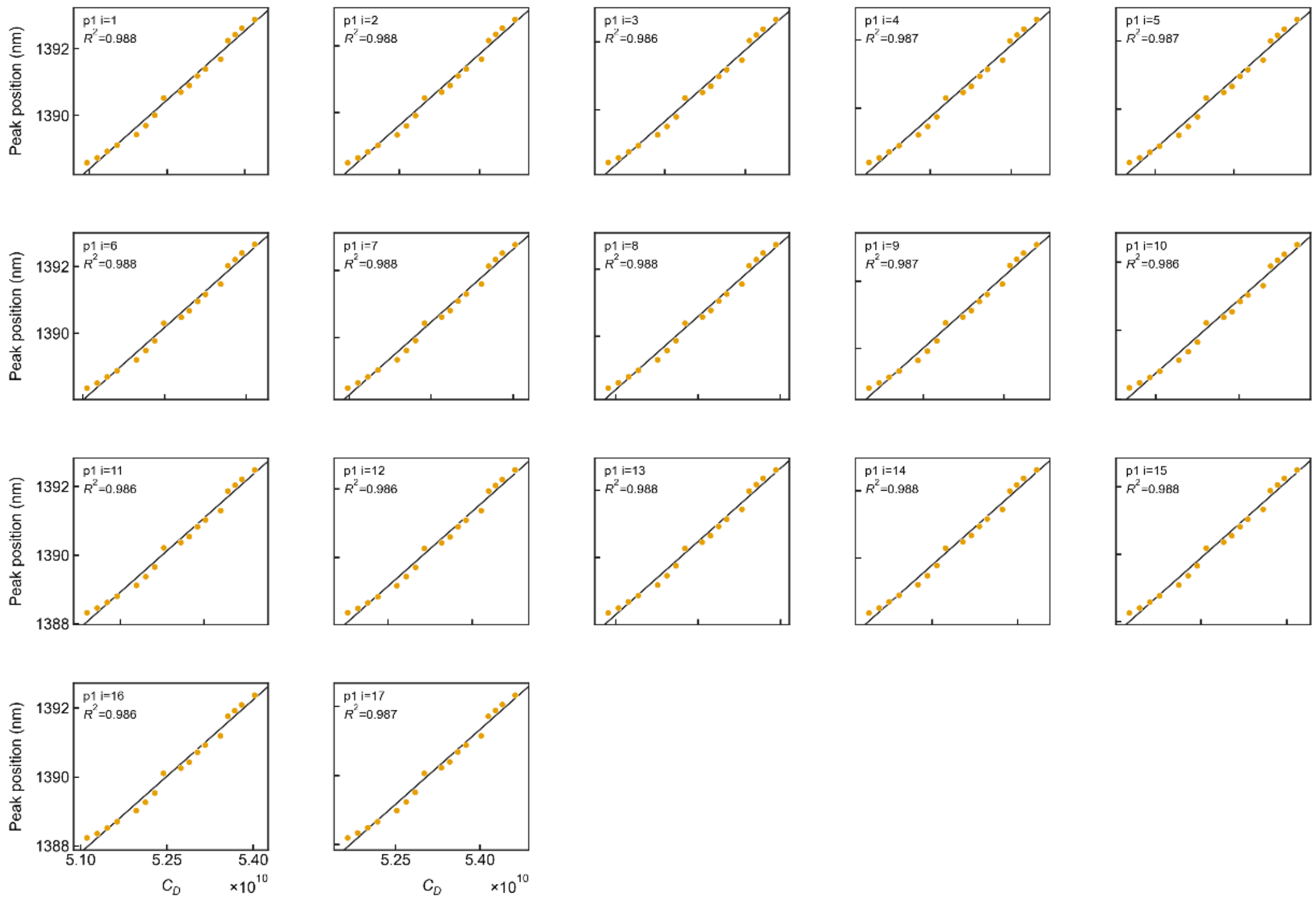

**Supplementary Fig. S8 | Resonance-position descriptor fits for all fixed-$P_1$ branches of AC.**

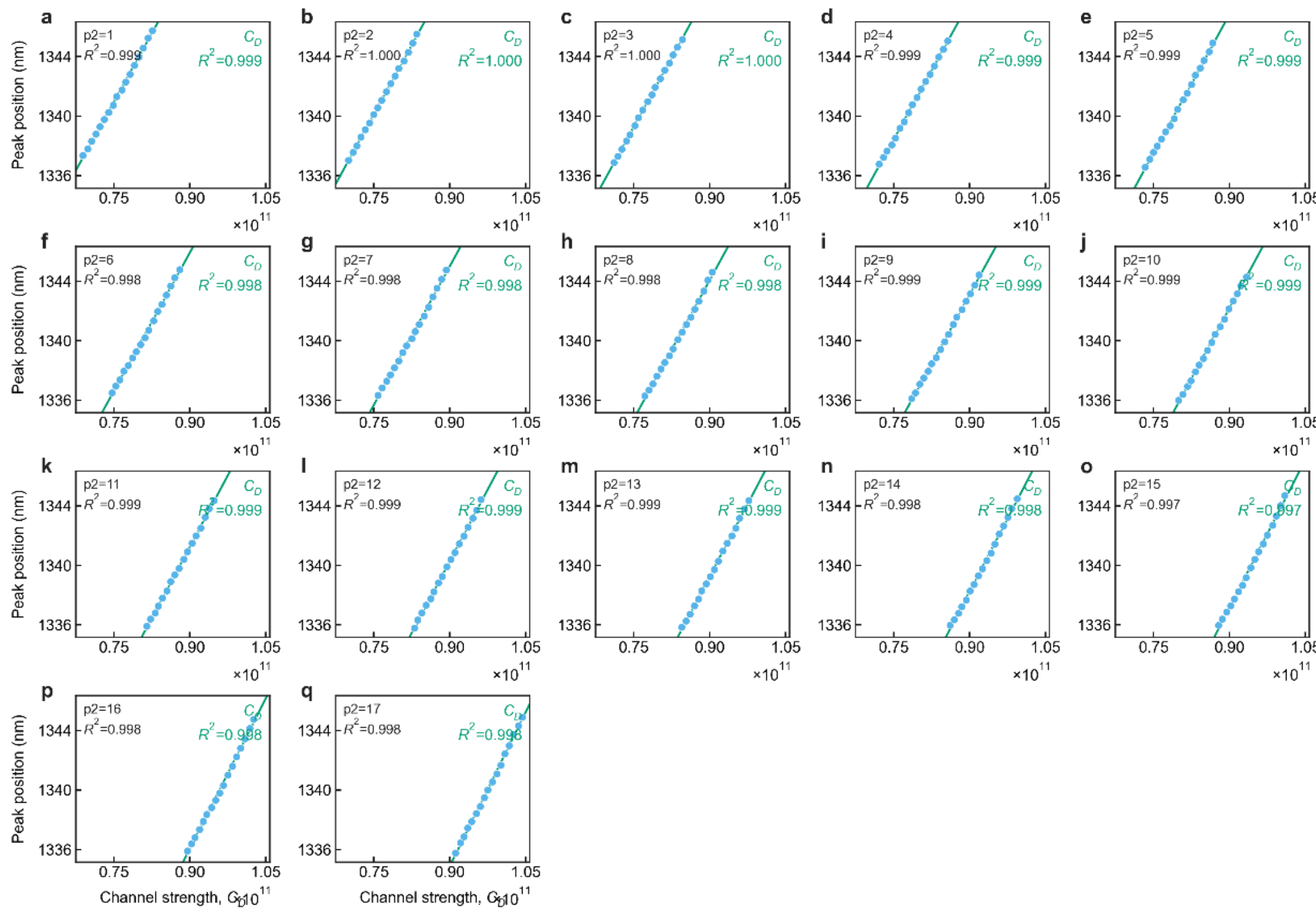

**Supplementary Fig. S9 | Resonance-position descriptor fits for all fixed-$P_2$ branches of AD.**

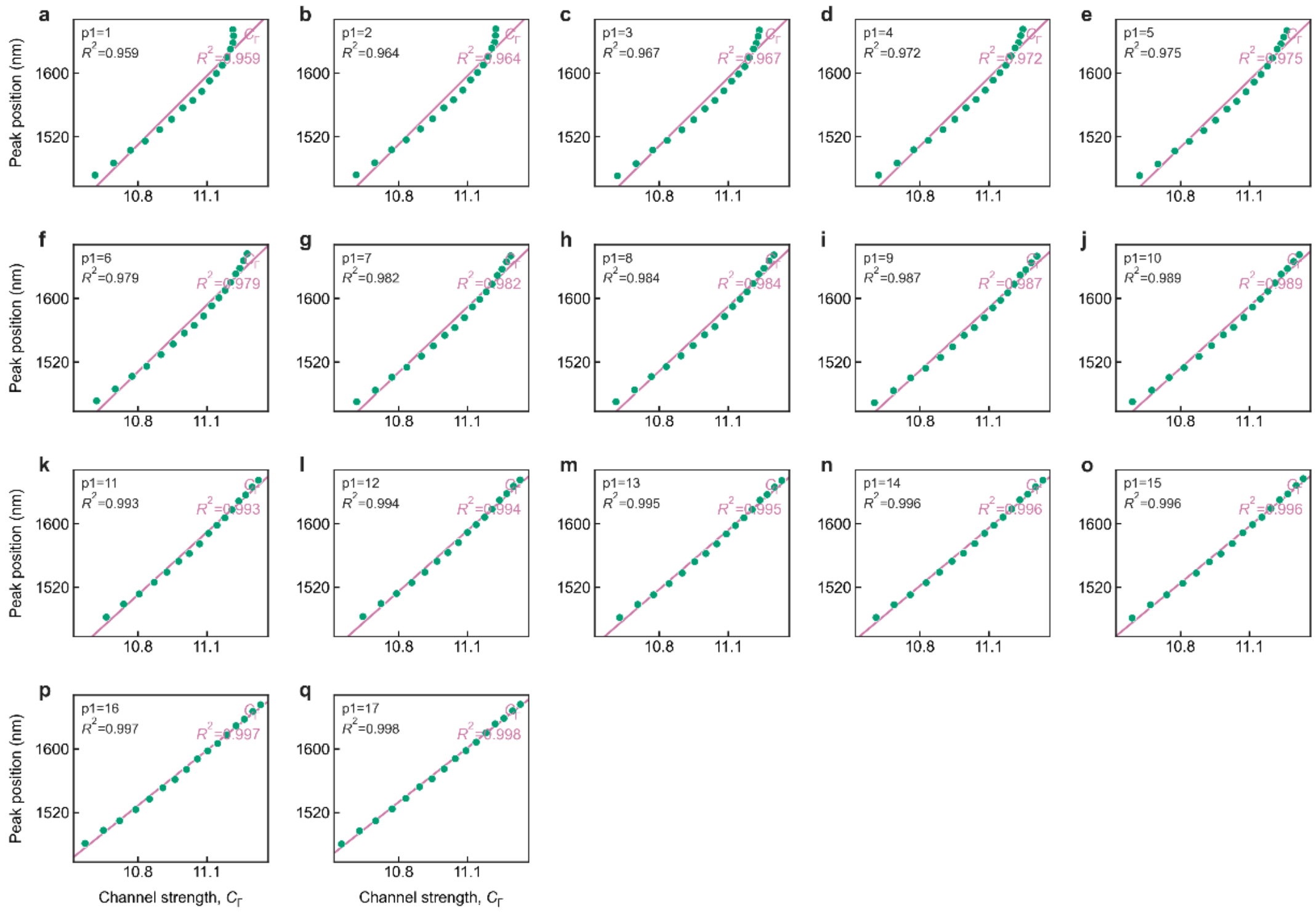

**Supplementary Fig. S10 | Resonance-position descriptor fits for all fixed-$P_1$ branches of CSR.**

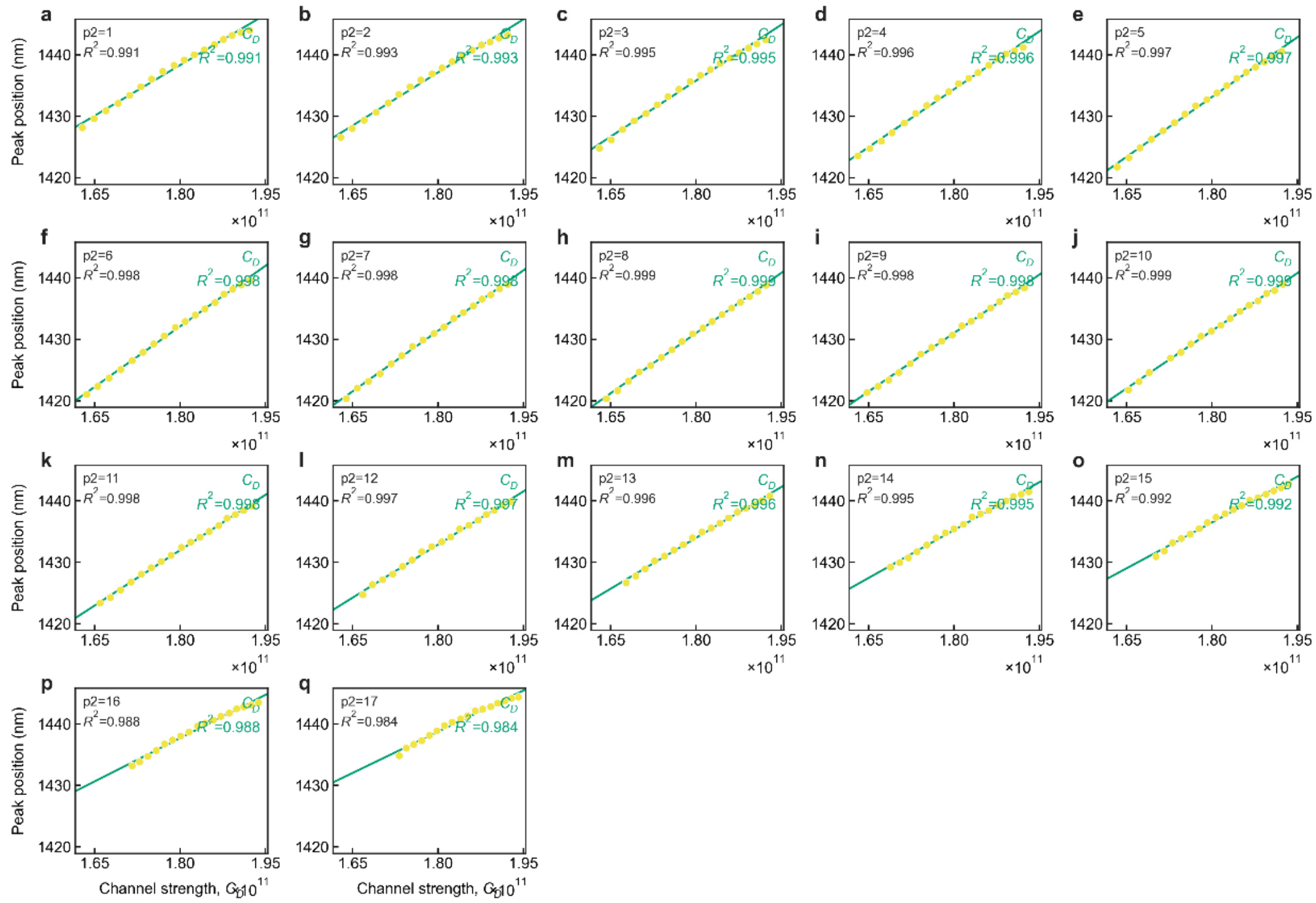

**Supplementary Fig. S11 | Resonance-position descriptor fits for all fixed-$P_2$ branches of DAN.**

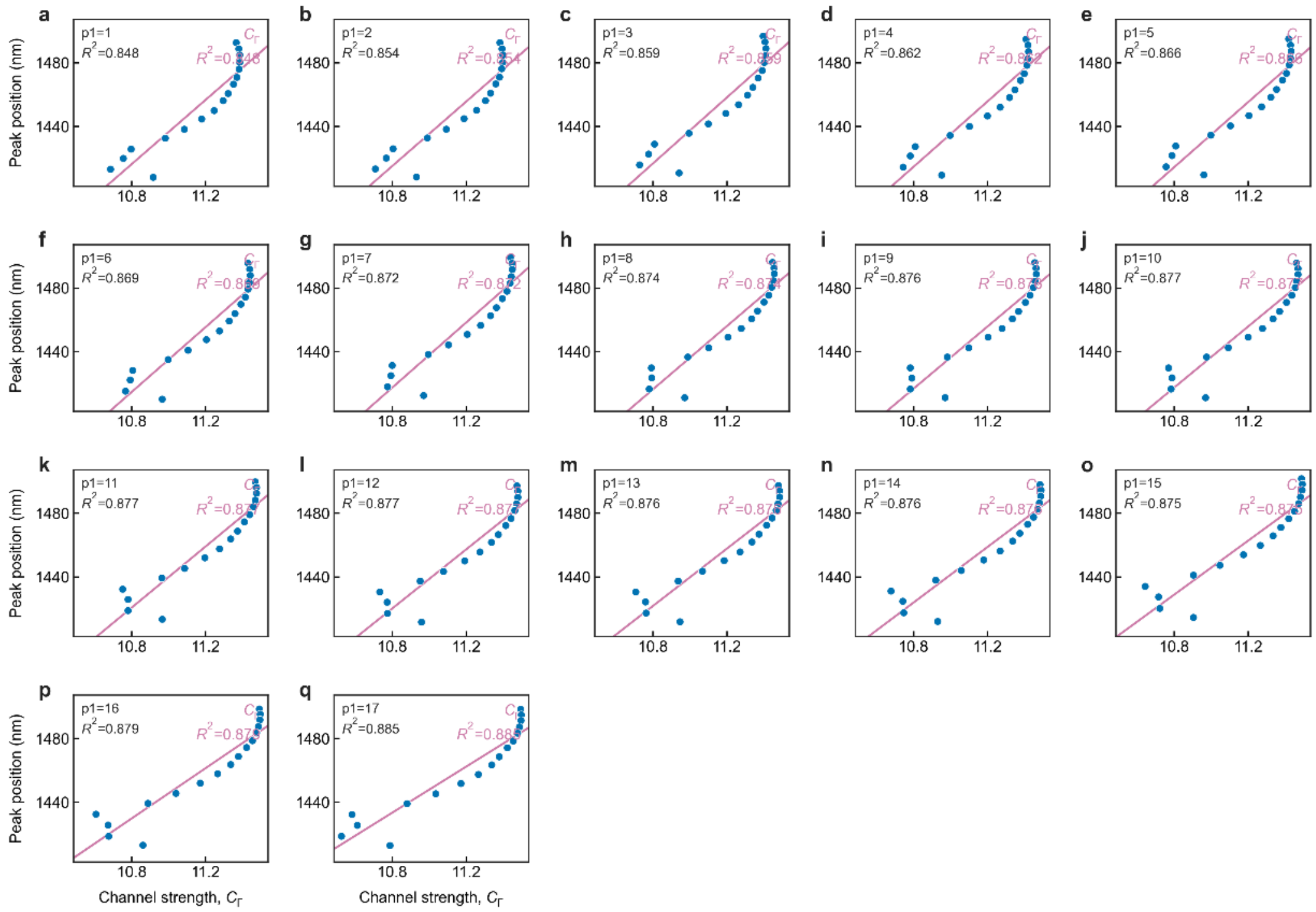

**Supplementary Fig. S12 | Resonance-position descriptor fits for all fixed-$P_1$ branches of PX.**

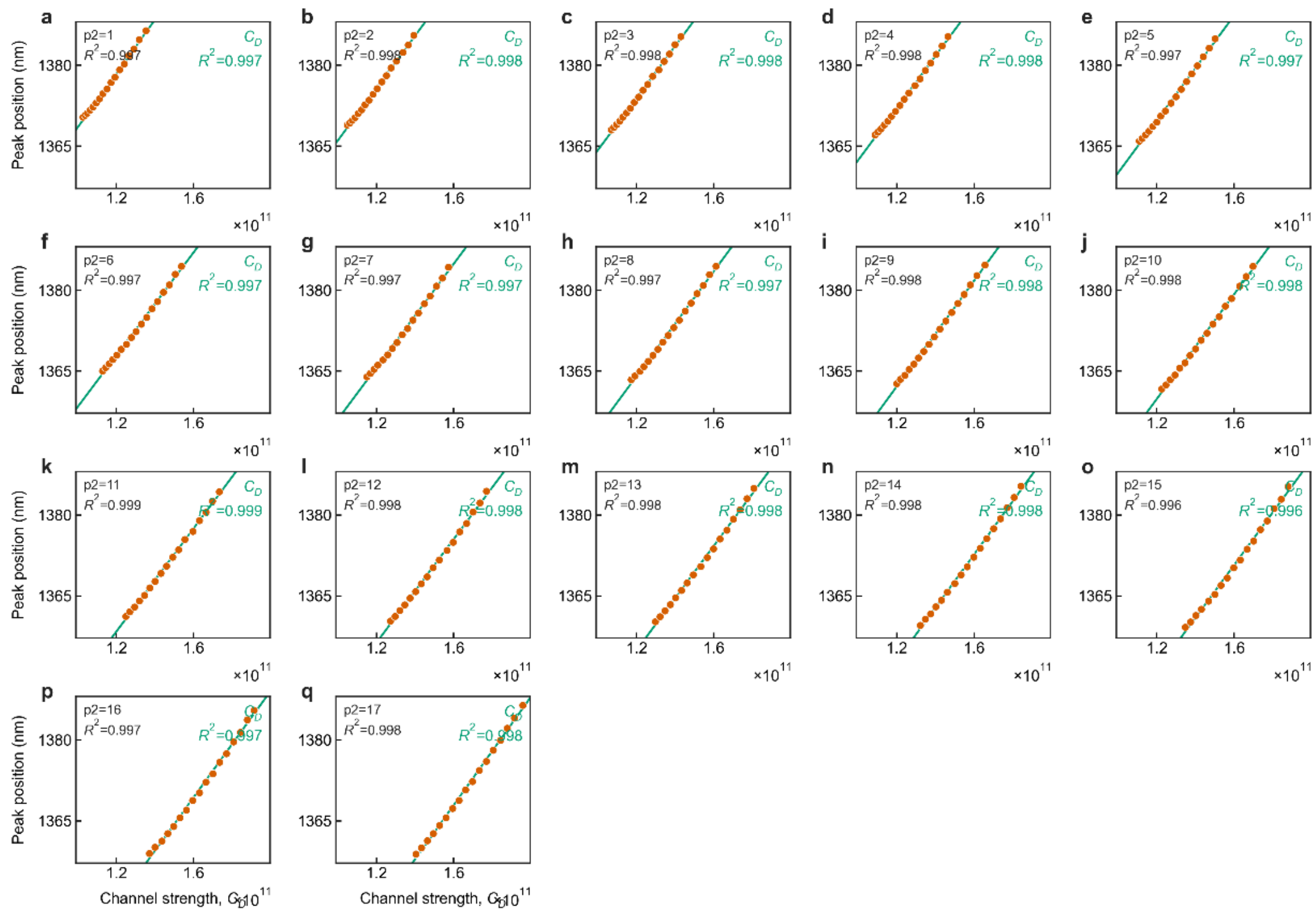

**Supplementary Fig. S13 | Resonance-position descriptor fits for all fixed-$P_2$ branches of SR.**

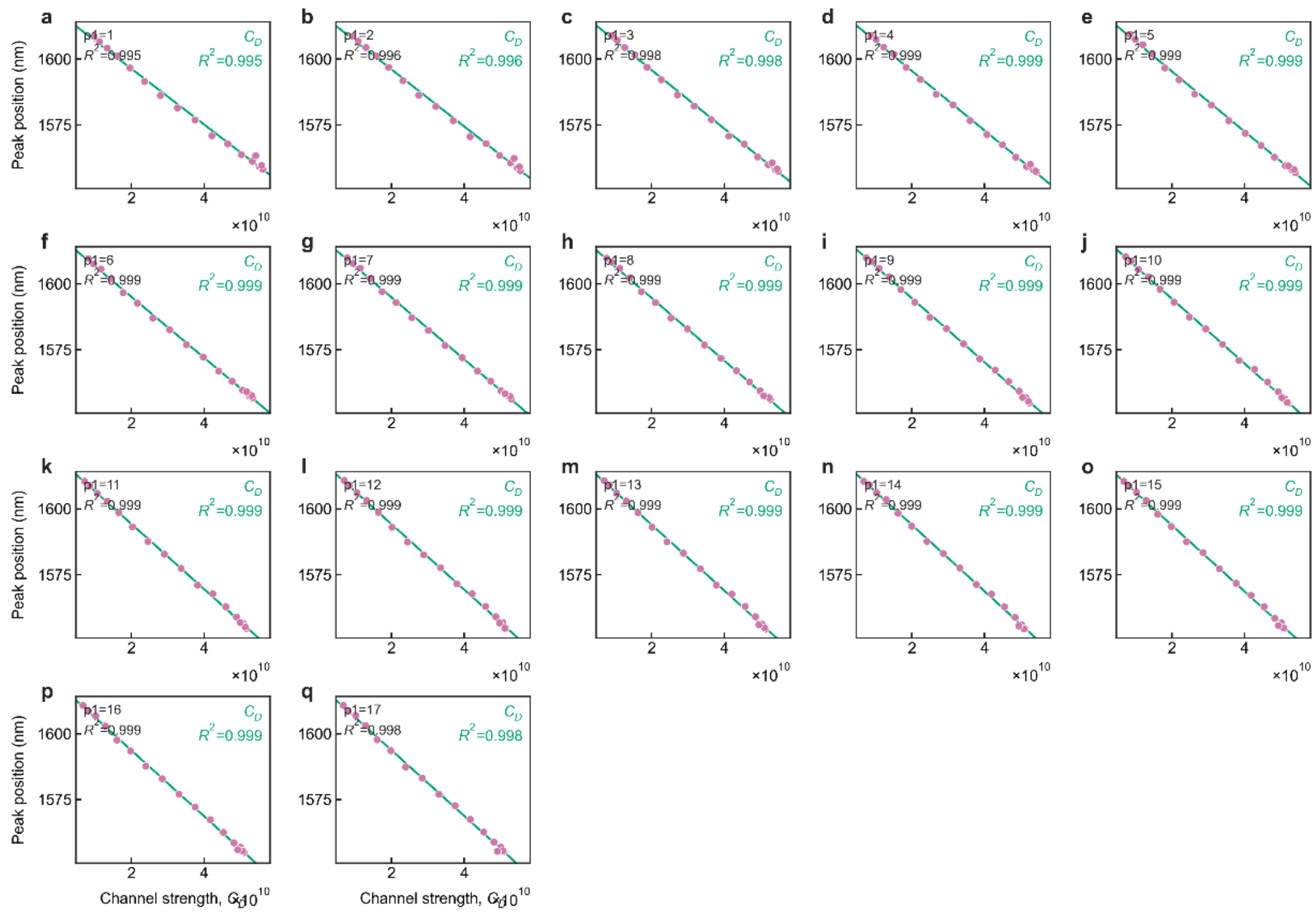

**Supplementary Fig. S14 | Resonance-position descriptor fits for all fixed-$P_1$ branches of TY.**

**Supplementary Note 4 | Field-level validation and reference-state dependence**

Complex driven fields at the reference resonance are used for TY2, TY4, TY8, CSR88, CSR90 and CSR93. An x–y FFT is performed independently at each z layer, and $|E_G|^2$ is summed over vector components and integrated over z to obtain the modal weight $M_G$. This field quantity differs from the dielectric coefficient $F_{mn}$. Figure S15 extends the low-order localization shown in main-text Fig. 4b to all six reference states.

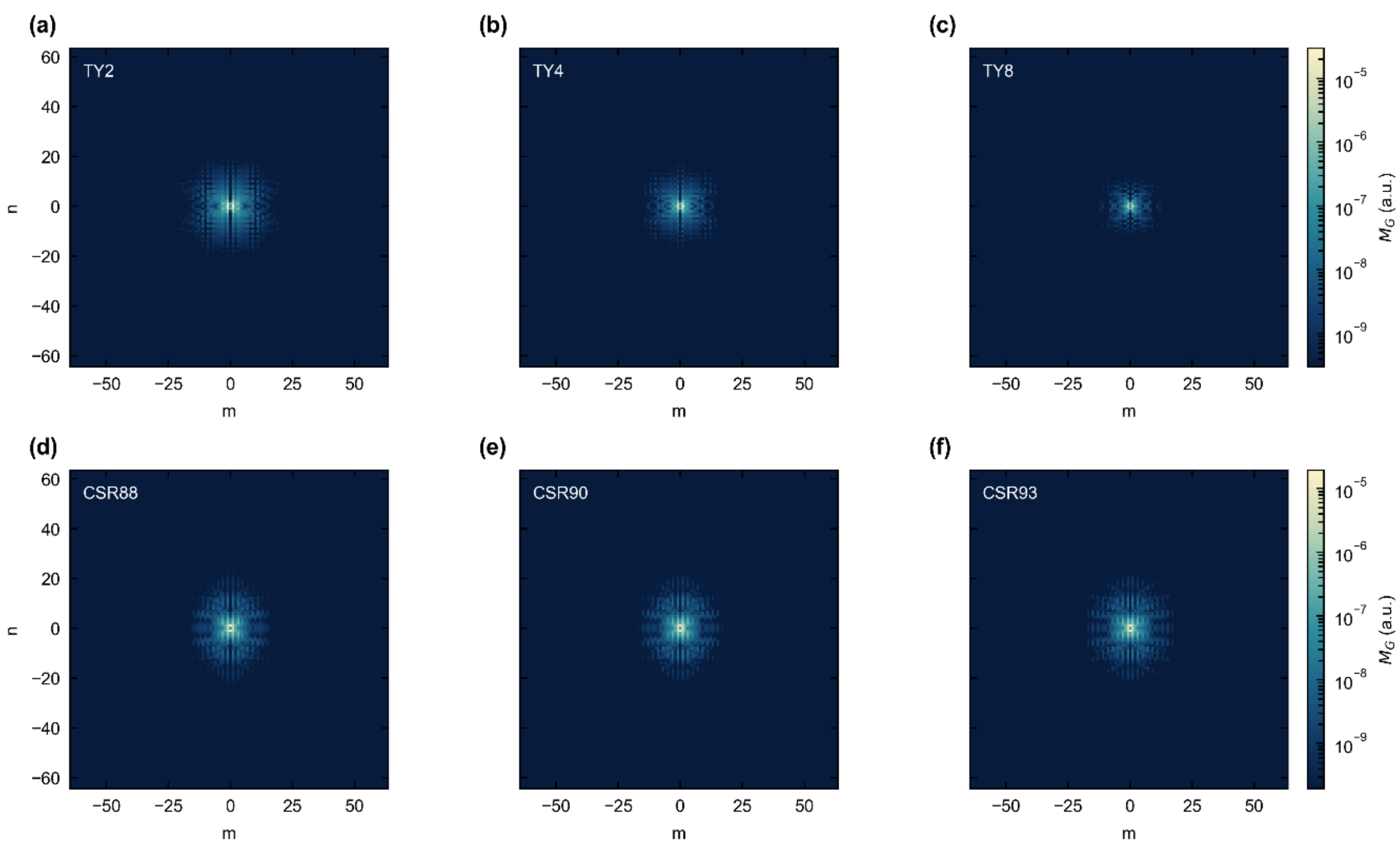

**Supplementary Fig. S15 | Modal Fourier weight $M_G$ of the six reference states.** The TY and CSR panels each share a logarithmic colour scale. Most of the weight lies in a finite low-order region near Γ, which supports substantial harmonic overlap at small nonzero **q**.

For each reference, the same 17-state branch is evaluated using the complete response $S_{all}$, the non-DC central 7 × 7 response $S_{low}$, and the directional $S_{C_D}$ for TY or $S_{C_\Gamma}$ for CSR. All S quantities retain complex $\Delta\varepsilon_q$ and $W_q$ weights; they are not the amplitude-only descriptors $C_\Gamma$ and $C_D$.

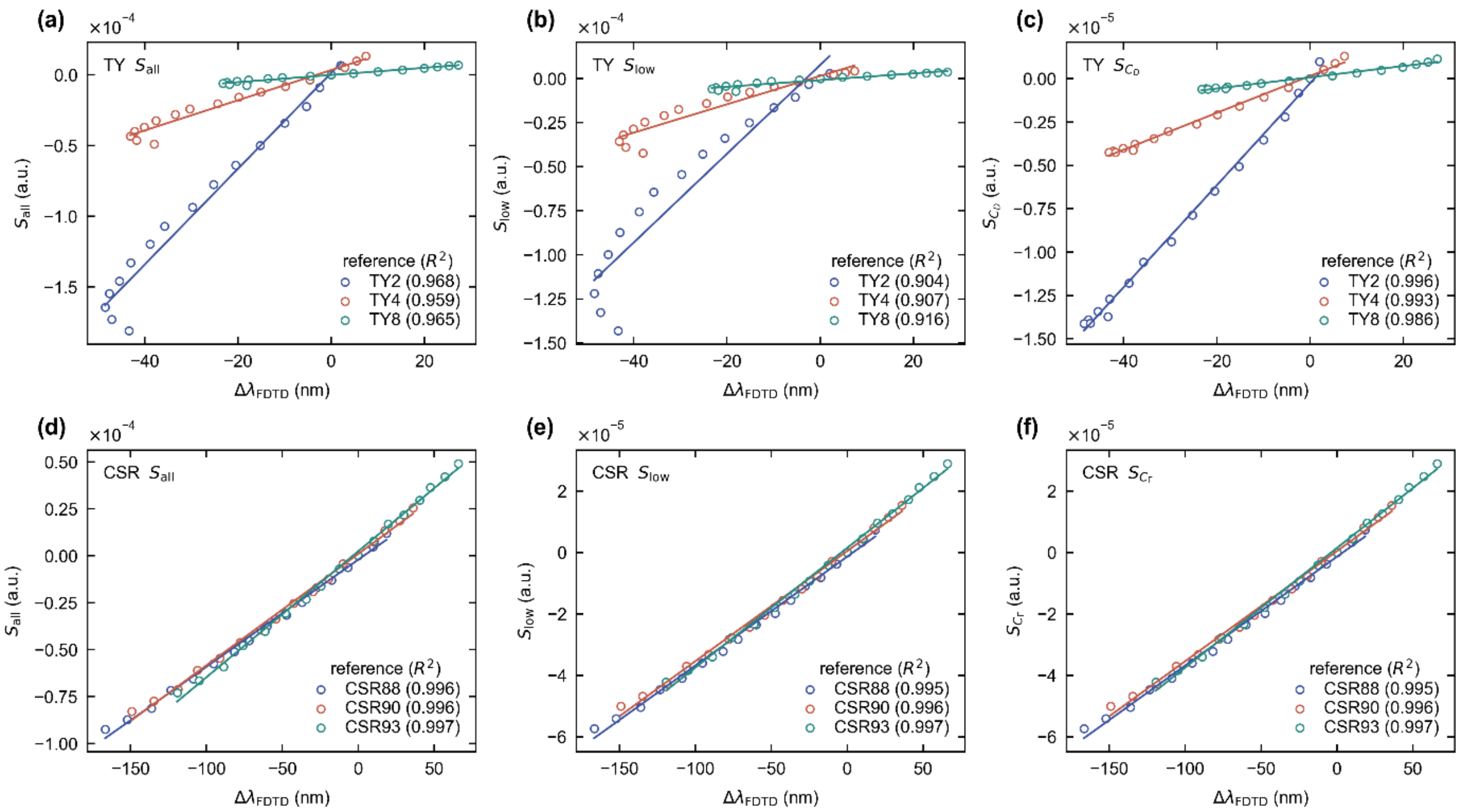

**Supplementary Fig. S16 | Partition of the field-overlap response.** a–c, $S_{\mathrm{all}}$, $S_{\mathrm{low}}$ and $S_{\mathrm{CD}}$ for TY; d–f, $S_{\mathrm{all}}$, $S_{\mathrm{low}}$ and $S_{\mathrm{C}\Gamma}$ for CSR. Values in parentheses are $R^2$ for each reference state; all panels use the same 17-state branch.

For TY, $R^2$ is 0.959–0.968 for $S_{all}$, 0.904–0.916 for $S_{low}$ and 0.986–0.996 for $S_{C_D}$. For CSR, $S_{all}$ gives 0.996–0.997 and $S_{low} = S_{C_\Gamma}$ gives 0.995–0.997. Main-text Fig. 4d uses TY2 with $S_{C_D}$ (0.996) and CSR88 with $S_{C_\Gamma}$ (0.995). $R^2$ measures linear trend agreement, not a captured fraction of frequency-shift energy. These driven-field tests support resonance-position trends; no normalized outgoing-channel projection is available here to validate $d_\mu$, $R_{\mu q}$ or radiative Q directly.

**Supplementary Note 5 | Forward learning, split boundaries and expert-input controls**

All four configurations use the complete 2,001-point window spectra, including the narrow resonance. Both backbones encode the full 67 × 67 input and its central 33 × 33 region into 32 PCA coefficients. The Fourier input has five channels: real part, imaginary part, log-amplitude, and sine and cosine of the phase. The expert receives only the central 5 × 9 × 9 block and predicts $\lambda_0$, ln $Q$, depth and Fano asymmetry $q_f$. Input statistics, family-dependent target statistics and the PCA basis were fitted on the training set only. Because no independent label exists for $q_f$, its target is set to zero, and it acts as a weakly regularized parameter learned through the spectral loss.

Training uses AdamW, batch size 40, weight decay $10^{-4}$ and gradient-norm clipping at 2.0. ReduceLROnPlateau uses a factor of 0.65 and patience 18. Background-only training has maximum/minimum epoch counts of 360/180, learning rate $8 \times 10^{-4}$ and early-stopping patience 90. Expert background, peak and joint stages use maximum/minimum epoch counts of 260/140, 360/180 and 420/220; learning rates $8 \times 10^{-4}$, $7 \times 10^{-4}$ and $1.5 \times 10^{-4}$; and patience 70, 90 and 110. Five seeds, 20260717–20260721, are evaluated after validation-based checkpoint selection. The background-only objective combines spectrum L1, 0.35 times spectrum MSE, 0.08 times standardized PCA-coefficient MSE and 0.08 times the squared second spectral difference. Expert optimization combines weighted spectrum L1 and 0.45 times weighted MSE, first- and second-difference errors weighted by 0.10 and 0.025, an output-range penalty weighted by 0.20, and standardized parameter supervision weighted by 1.0 during peak training or 0.80 during joint training. Local training weights use the labeled center; the true center, Q and depth are not supplied at inference.

The strict split separates source-grid geometry hashes and principal-scan parameter blocks (1,556/238/221 samples). At the 67 × 67 binary-mask resolution, 68 test masks nevertheless coincide with a training mask, because distinct source geometries can round to the same mask. The other 153 form the no-training-repeat subset, which still belongs to the seven known families. Figure S17 summarizes this distinction; random-split counts are shown for reference only.

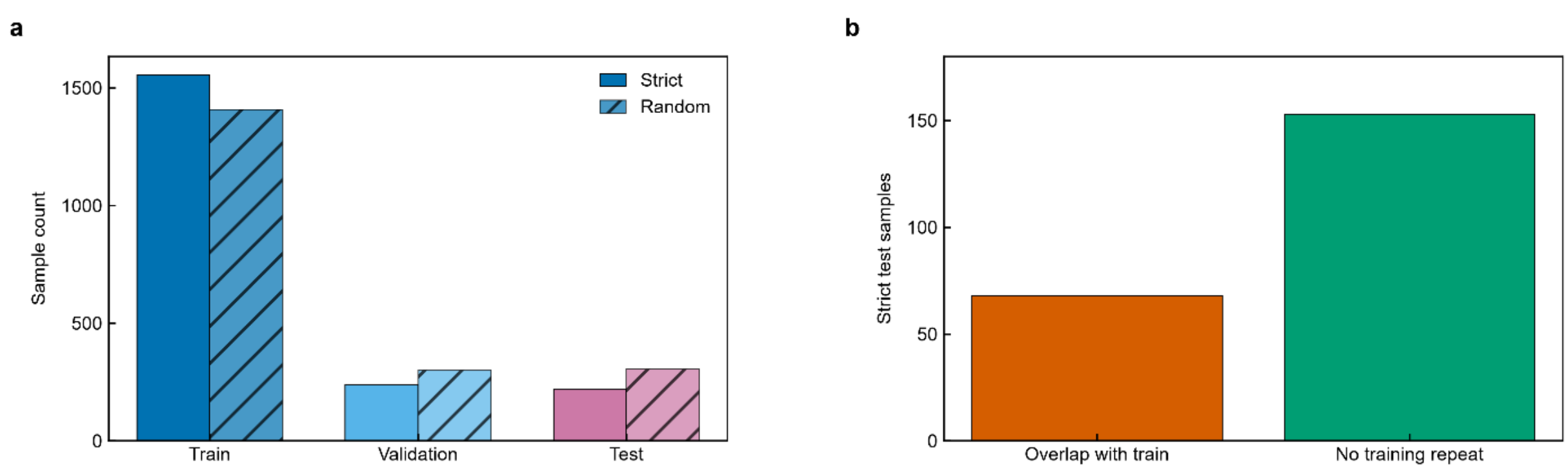

**Supplementary Fig. S17 | Strict split, random split, and geometric duplication.** a, Training, validation, and test sample counts under the two splits. b, Partition of the strict test set into 68 samples whose binary geometries are exactly duplicated in the training set and 153 samples with nonduplicated geometries. Novel geometry denotes a test geometry not duplicated in training but still belonging to one of the seven known families within the current scan range.

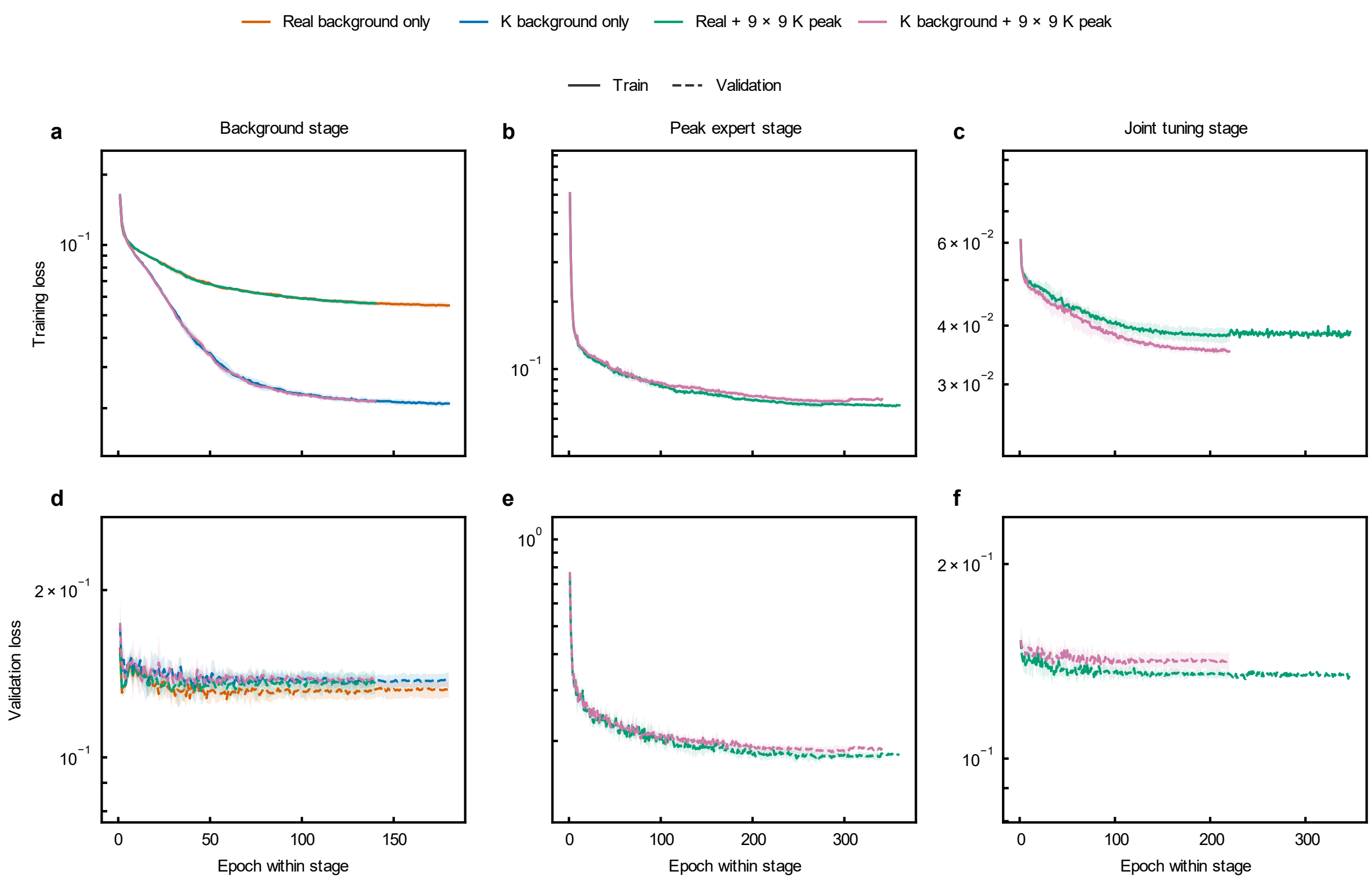

**Supplementary Fig. S18 | Training and validation losses of the four forward models.** a–c, Training losses during background, peak-expert and joint-tuning stages. d–f, Corresponding validation losses. Colors identify the four models; solid and dashed lines denote training and validation. Lines and shading show the mean and one sample standard deviation across available histories from five seeds (20260717–20260721). Early stopping gives different curve lengths and may reduce the contributing seed count at later epochs. Background-only models have no peak-expert or joint-tuning stage. Epochs restart at each stage. Loss axes are logarithmic; objectives differ between stages.

Complete-spectrum MAE averages absolute error over the target window. Local errors are evaluated within ±8 or ±5 nm of the labeled center, and core error within ±0.5 nm; these masks are evaluation definitions, not inference inputs. Minimum-transmission and resonance-position MAEs quantify the dip depth and position, while median spectral R summarizes samplewise Pearson correlation. Figure S19 adds ±5 nm and core metrics to main-text Table 1; bars summarize five seeds and error bars are sample standard deviations.

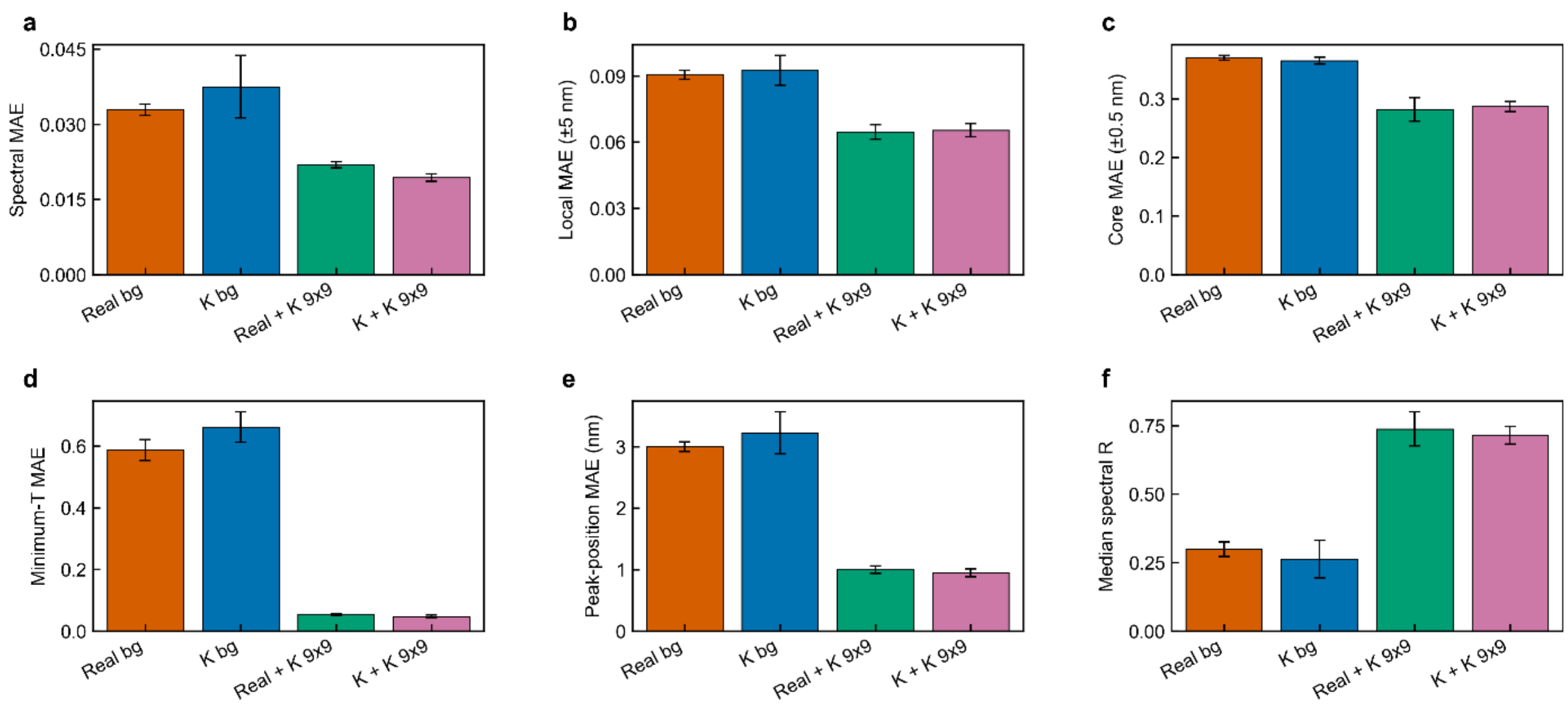

**Supplementary Fig. S19 | Five-seed test metrics for the four forward models under the strict split.** a–f, Complete-spectrum MAE, ±5 nm local MAE, ±0.5 nm core MAE, minimum-transmission MAE, resonance-position MAE, and median spectral R, respectively. Bar heights are five-seed means, error bars are sample standard deviations, and the test set contains 221 samples.

The gain is not uniform across families (Fig. S20). The expert reduces the resonance-position error for AD, CSR, DAN, PX and SR. AC errors are already small and change little, and for TY the expert increases both the mean error and its spread across seeds.

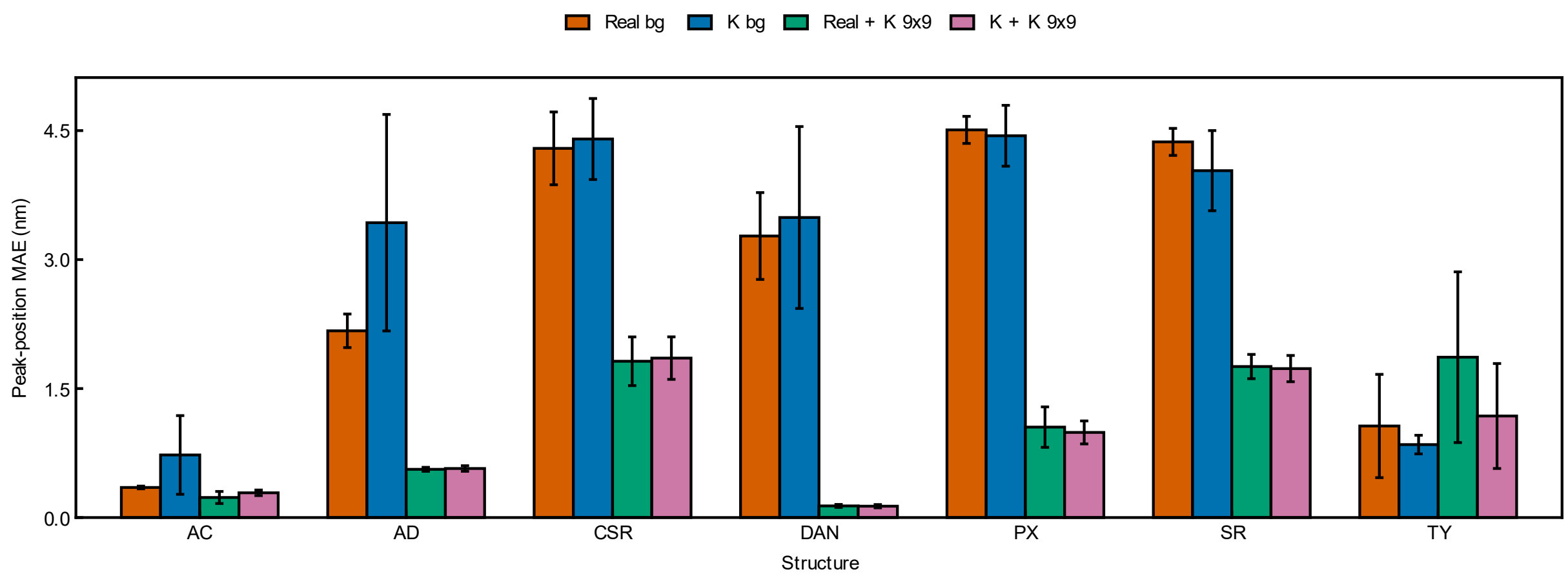

**Supplementary Fig. S20 | Family-resolved resonance-position MAE.** Real bg, K bg, Real + K peak and K + K peak are compared for each family. Bars show five-seed means on the strict test set; error bars, s.d.

The input ablation keeps the real-space backbone, split and seed (20260717) fixed and varies only the expert input: the central complex Fourier block (9 × 9), the full real-space map (67 × 67), the full Fourier tensor (67 × 67), a central 9 × 9 real-space crop and the scalar pair ($C_D$, $C_\Gamma$). The resonance-position MAEs are 0.996, 2.049, 1.979, 2.691 and 2.548 nm, and the minimum-transmission MAEs are 0.052, 0.136, 0.144, 0.327 and 0.235. This single-seed control distinguishes the representations but does not rank them statistically.

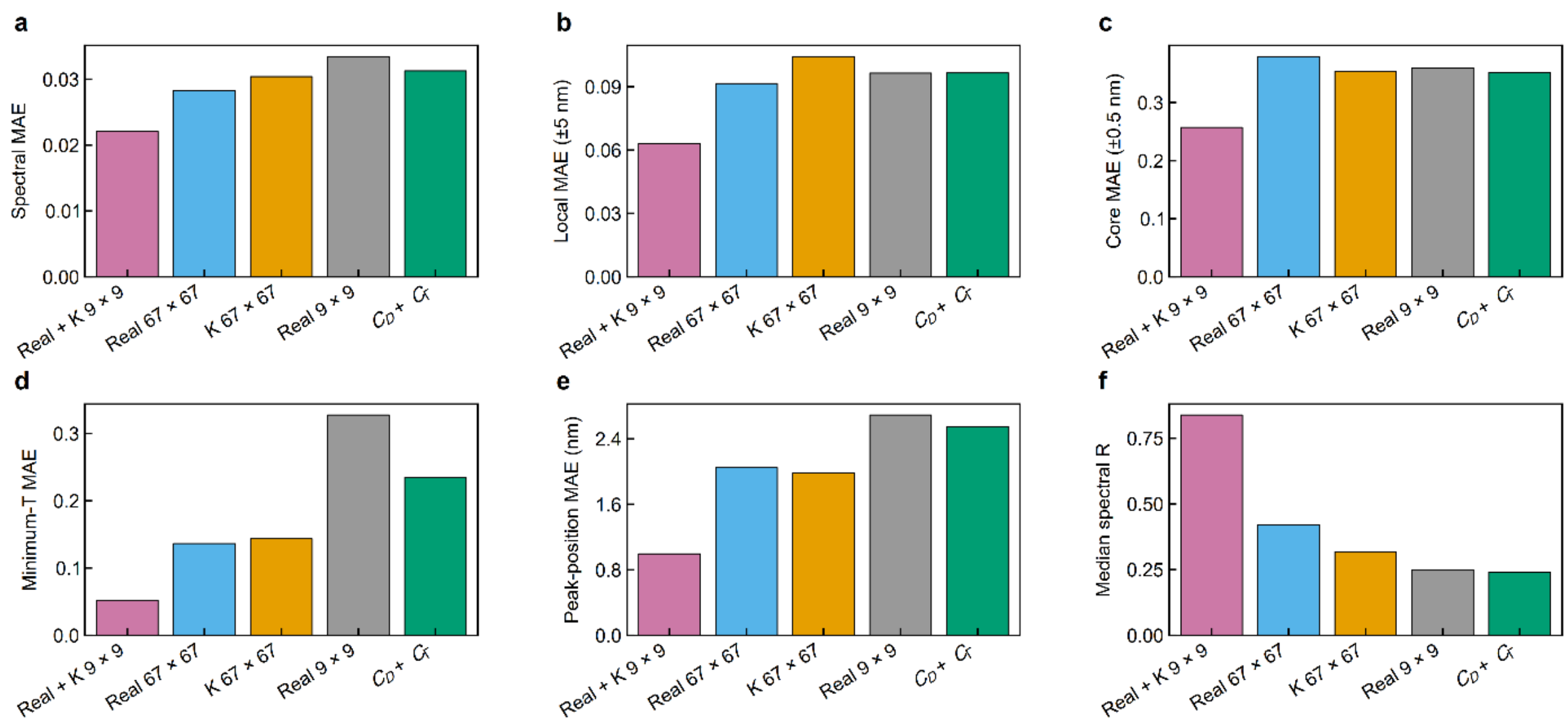

**Supplementary Fig. S21 | Expert-input ablation.** The five inputs are the central 9 × 9 complex Fourier block (Real + K 9 × 9), the full real-space map (Real 67), the full Fourier tensor (K 67), a central 9 × 9 real-space crop (Real 9 × 9) and the scalar pair $C_D + C_\Gamma$. a–f, Complete-spectrum MAE, ±5 nm local MAE, ±0.5 nm core MAE, minimum-transmission MAE, resonance-position MAE and median spectral $R$. Single seed, strict split.

Additional surrogate-model controls use seed 20260717, the same strict split (1,556/238/221) and all 2,001 target points. Real-space ResNet and a two-dimensional Fourier neural operator (FNO) directly regress spectra; an adapted offline two-stage model regresses training-only PCA32 background coefficients and independently fitted single-resonance Fano parameters, followed by analytical synthesis. The latter is a Ma-style adaptation, rather than an exact reproduction of RIDL. All models receive the same family information.

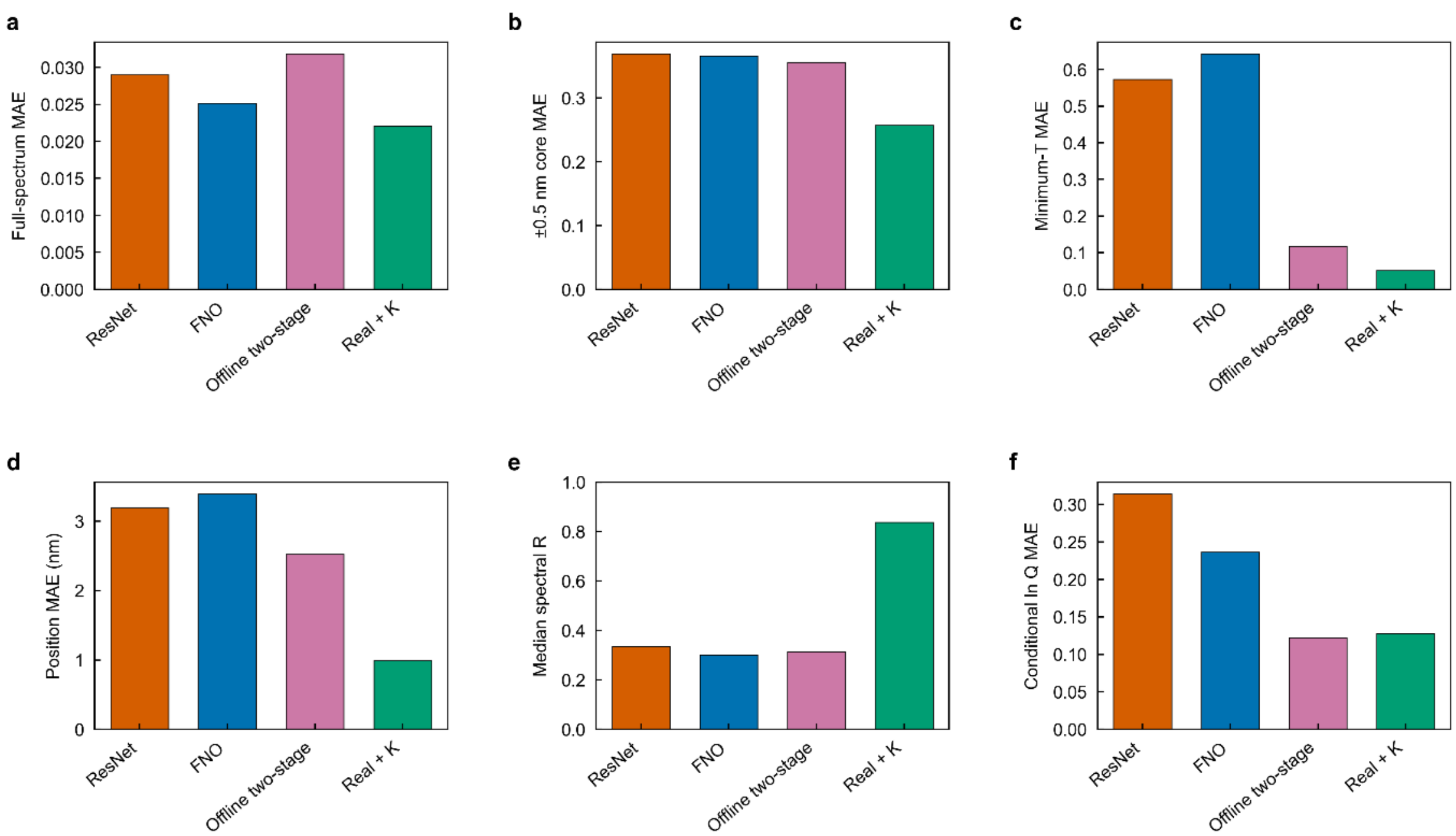

**Supplementary Fig. S22 | Additional surrogate-model comparison.** a–f, Complete-spectrum MAE, ±0.5 nm core MAE, minimum-transmission MAE, resonance-position MAE, median spectral Pearson R and conditional ln Q MAE. All models use the same 221 strict-test samples and seed 20260717. The ln Q error is conditional on both spectra being resolvable; each model uses its respective resolvable intersection. Bars are single-seed results without uncertainty estimates.

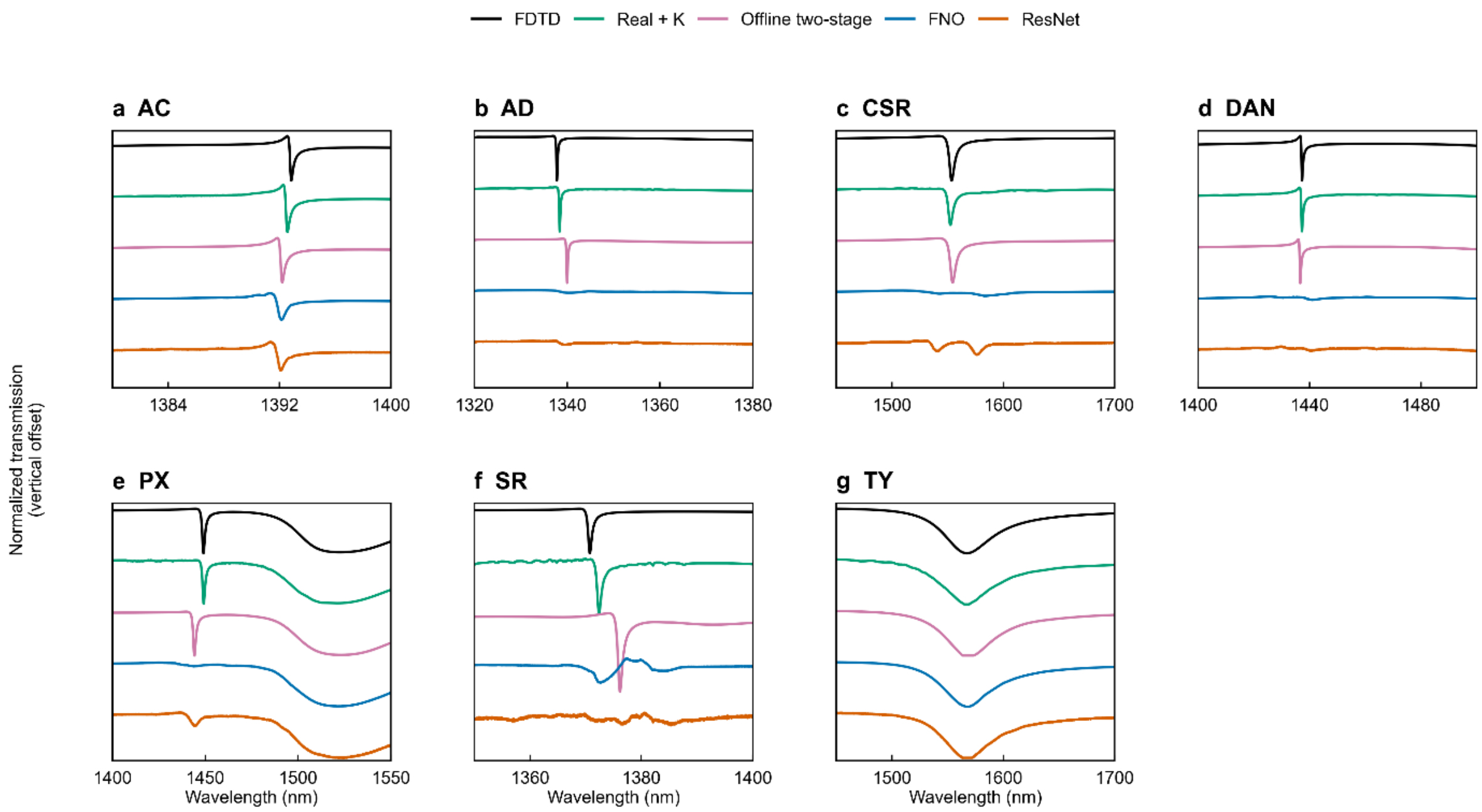

**Supplementary Fig. S23 | Representative offset spectra from the additional surrogate models.** a–g, AC, AD, CSR, DAN, PX, SR and TY. Each family uses the middle sample after lexicographically sorting its strict-test sample IDs, independently of prediction error. Within each sample, all curves share the normalization (T − min T_FDTD)/(max T_FDTD − min T_FDTD). From top to bottom, FDTD, Real + K, offline two-stage, FNO and ResNet are shifted vertically by 4.60, 3.45, 2.30, 1.15 and 0, respectively. Offsets are for display only; every curve retains all 2,001 wavelength points.

The 2,015 effective FDTD simulations required 146,984.4 CPU·h in total, corresponding to 72.95 CPU·h and 2.28 wall-clock h per sample on 32 allocated CPU cores. GPU training times include validation and checkpoint overhead but exclude preprocessing. On an NVIDIA GeForce RTX 4050 Laptop GPU, inference with prepared 67 × 67 inputs at batch size one requires 1.593–3.403 ms per sample, excluding geometry construction, FFT, disk I/O and model loading; transfer-inclusive timings are reported separately. CPU·h and GPU·h are reported independently rather than converted into a direct speedup.

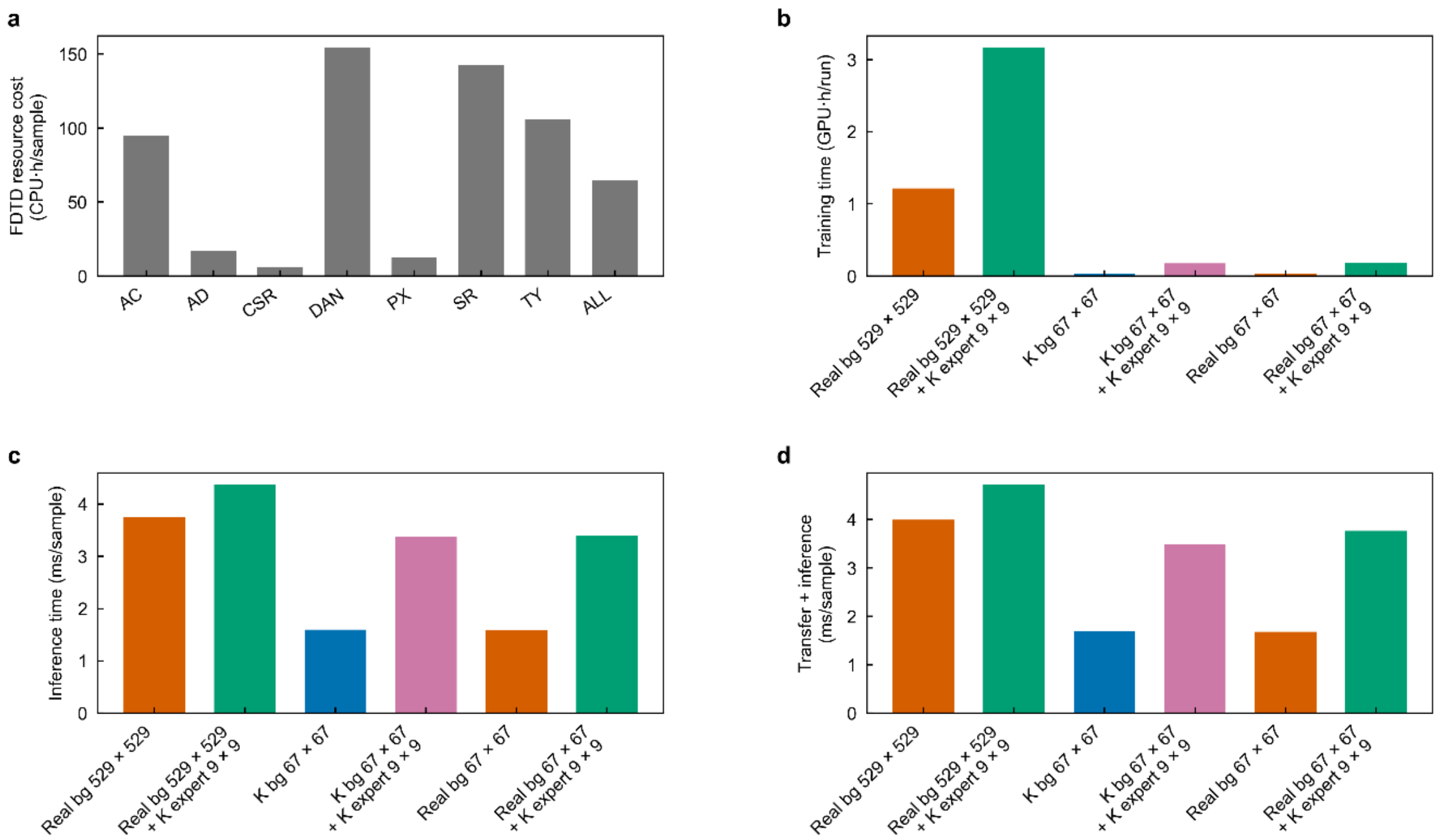

**Supplementary Fig. S24 | FDTD, training and inference time.** a, Median FDTD CPU resource cost per sample. b, Recorded training GPU time per run. c, Prepared-input, device-resident model inference at batch size one. d, Prepared-input transfer, inference and output transfer. The labels explicitly distinguish background-network inputs (Real or K, 529 × 529 or 67 × 67) from the Fourier peak-expert input (K, 9 × 9). Background-only models have no peak expert. The 67 × 67 background branches show means across five seeds; the 529 × 529 background branches have one recorded seed. No error bars are displayed.

## Supplementary Note 6 | Leave-one-family-out and one-sample adaptation

We evaluated the K + K model on seven held-out families: AC, AD, CSR, DAN, PX, SR and TY. For each family, the zero-sample setting excluded that family from training and validation. In the one-sample setting, one labeled spectrum from its strict-split training pool was added to training. Adding one sample reduced full-spectrum MAE for all seven families. The change in resonance-position MAE depended on the family (Fig. S25).

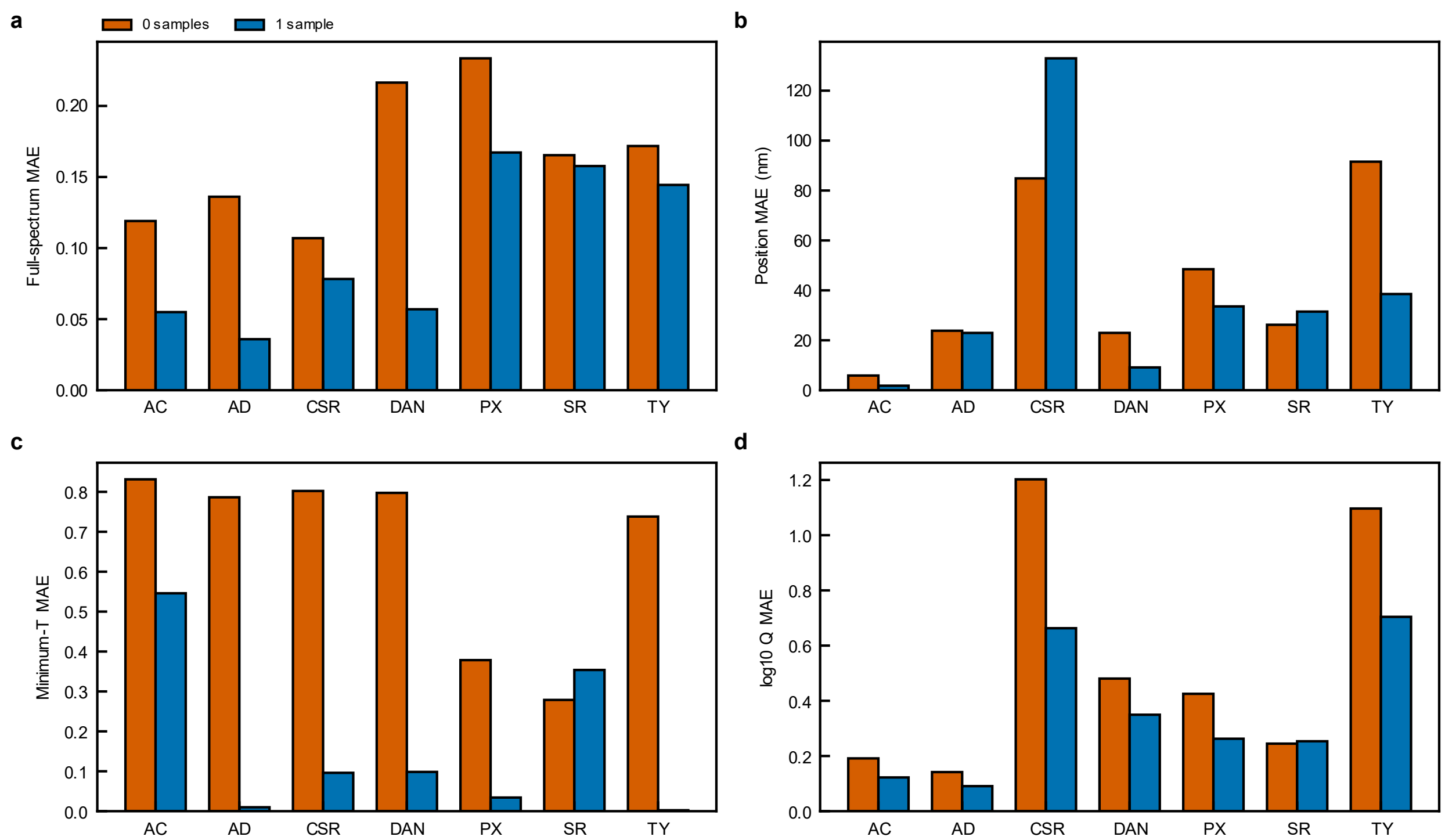

Supplementary Fig. S25 | Leave-one-family-out and one-sample adaptation of the K + K model. a–d, Full-spectrum MAE, resonance-position MAE, minimum-transmission MAE and log10 Q MAE for AC, AD, CSR, DAN, PX, SR and TY. Orange and blue bars denote zero and one labeled sample from the held-out family, respectively. Both settings use the same fixed test samples for each family and seed 20260717. Resonance position is the wavelength of minimum transmission within the BIC window. The log10 Q MAE compares the model-predicted log10 Q with the interpolated Q label. Bars show one seed and one selected support sample per family.

**Supplementary Note 7 | Inverse reconstruction and its evaluation boundary**

The inverse encoder receives the 2,001-point BIC spectrum and normalized wavelength. Seven-family labels provide auxiliary training supervision; only predicted class probabilities enter the decoder at inference. The target is a binary 67 × 67 material mask. In the Fourier routes, mask logits are set to +4 for foreground and −4 for background, then transformed by orthonormal rFFT2 and standardized using training-only coefficient means and scales. The decoder predicts the complex coefficients, reverses this standardization and applies irFFT2, sigmoid and a 0.5 threshold. This target is the Fourier representation of mask logits, not the mean-subtracted physical dielectric coefficients used in descriptor analysis.

The direct-real objective is weighted binary cross-entropy + 0.75 Dice loss + 0.20 auxiliary class cross-entropy. Joint supervision adds 0.08 SmoothL1 loss on standardized Fourier coefficients. K-only uses SmoothL1 on these coefficients + 0.20 class cross-entropy, without real-space BCE or Dice. Training uses AdamW at $7 \times 10^{-4}$, weight decay $10^{-4}$, batch size 48, a maximum of 220 epochs and early-stopping patience 35; ReduceLROnPlateau uses a factor of 0.5 and patience 10. This inverse comparison uses a single seed.

Figure S26 reports results for all 221 test samples and for the 153 masks absent from training. On the full test set, IoU is 0.944 (Fourier only), 0.926 (joint) and 0.853 (direct real), and probability MAE is 0.030, 0.015 and 0.027. On the no-training-repeat subset, IoU is 0.927, 0.920 and 0.816. IoU measures the thresholded contour, probability MAE the calibration of the unthresholded map, and exact-mask accuracy requires every pixel to match; the three are complementary. The nonzero exact-mask accuracy of the Fourier-only route comes from the duplicated masks (Fig. S26f).

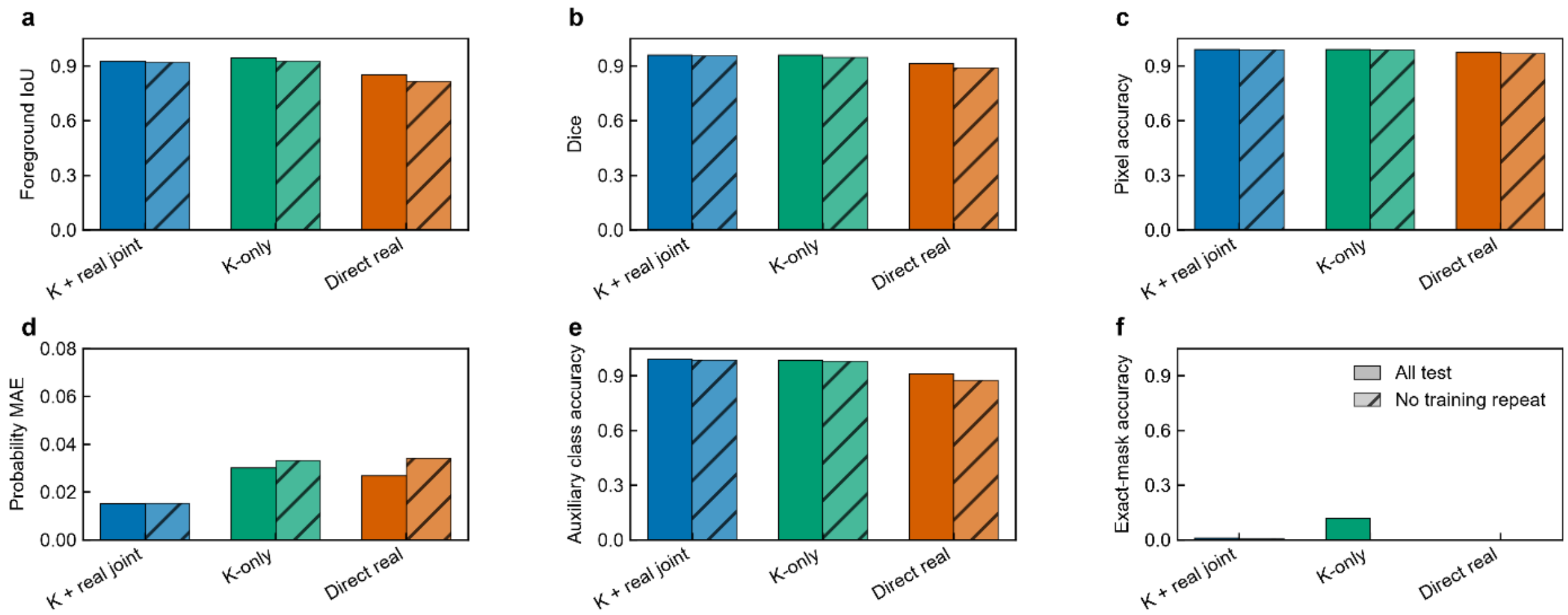

**Supplementary Fig. S26 | Test metrics for the three inverse-reconstruction routes.** a–f, Foreground IoU, Dice score, pixel accuracy, probability MAE, auxiliary structure-family accuracy, and exact-mask accuracy, respectively. Solid bars denote the complete test set (n = 221), and hatched bars denote the novel-geometry subset (n = 153).

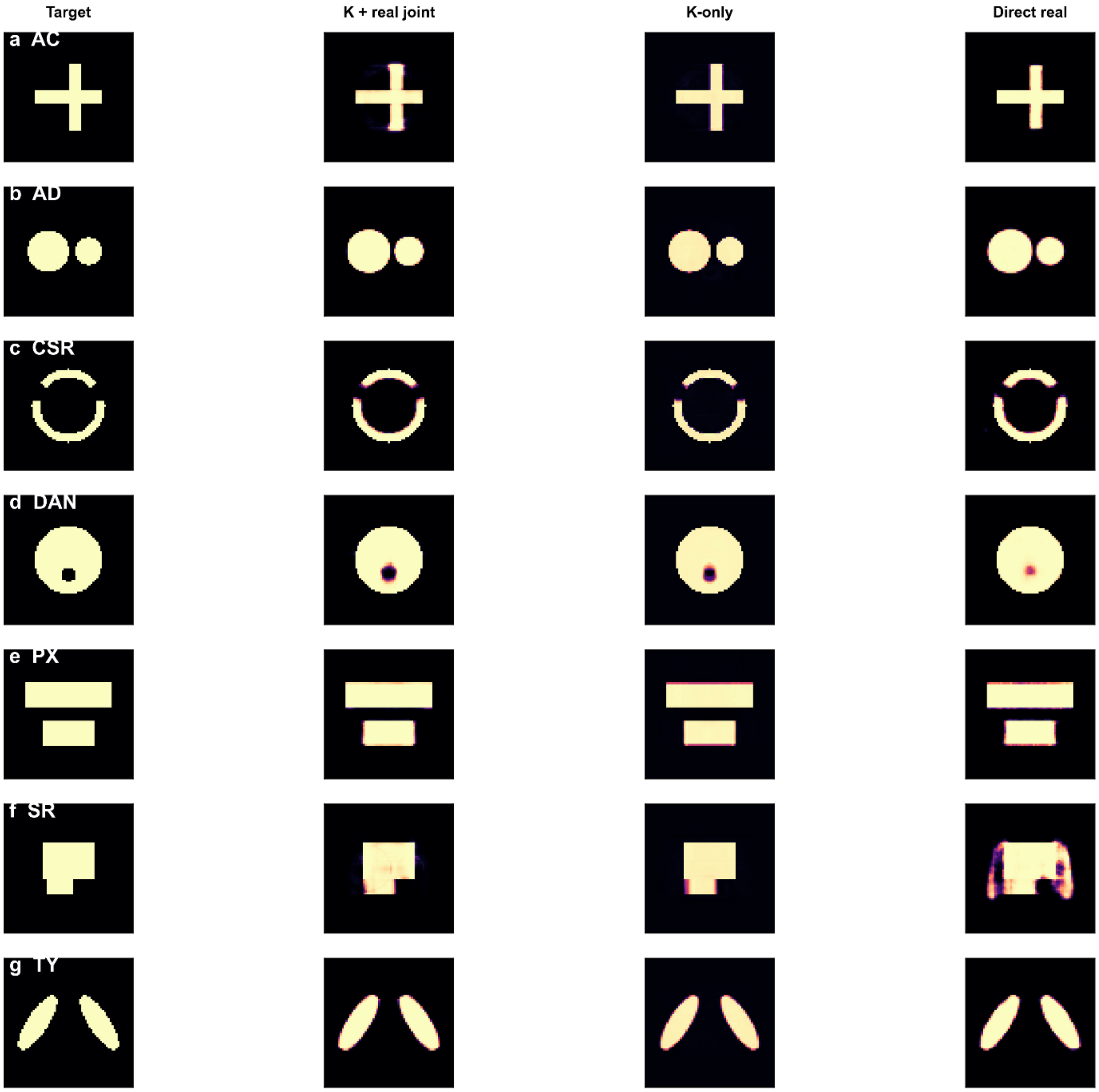

**Supplementary Fig. S27 | Inverse reconstruction of representative test samples from the seven structure families.** Each row corresponds to one family; columns compare the ground-truth dielectric distribution with K + real joint, K-only, and Direct real. Predicted maps show unthresholded foreground probabilities from 0 to 1.

The seven-family examples in Fig. S27 extend main-text Fig. 5b. Reconstruction is conditional on the paired dataset and known topology classes; spectral nonuniqueness and generalization to unseen topologies remain unresolved.